\documentclass{emulateapj}
\usepackage{lscape}
\usepackage{apjfonts}

\newcommand{\Msol}{M$_{\odot}$}

\newcommand{\Mjup}{M$_{\mathrm{JUP}}$}

\newcommand{\kms}{km\,s$^{-1}$}

\newcommand{\simgt}{\gtrsim}

\newcommand{\simlt}{\lesssim}

\newcommand{\Ms}{CH$_{\mathrm 4}$s}
\newcommand{\Ml}{CH$_{\mathrm 4}$l}
\newcommand{\Msl}{\Ms$-$\Ml}

\slugcomment{To appear in  {\em The Astronomical Journal}, Nov. 2005}
\shortauthors{Tinney et al.}
\shorttitle{Hunting out T dwarfs with Methane Imaging}
\received{2005 February 19}
\accepted{2005 July 21}
\begin{document}

\title{The 2MASS Wide-Field T Dwarf Search. IV. Hunting out T dwarfs with Methane Imaging\altaffilmark{1}}
\author{C. G. Tinney\altaffilmark{2}, 
        Adam  J. Burgasser\altaffilmark{3,6}, 
        J. Davy. Kirkpatrick\altaffilmark{4},
        and Michael W. McElwain\altaffilmark{5}
        }
\altaffiltext{1}{Based on observations obtained at the 
Anglo-Australian Telescope, Siding Spring. Australia.}
\altaffiltext{2}{Anglo-Australian Observatory, PO Box 296. 
                Epping. 1710 Australia. {\tt cgt@aaoepp.aao.gov.au}}
\altaffiltext{3}{Department of Astrophysics, American Museum of Natural History, New York NY 10024 USA. {\tt adam@amnh.org}.}
\altaffiltext{4}{Infrared Processing \& Analysis Center, Caltech, Pasadena, CA. 91125  USA. {\tt davy@ipac.caltech.edu}}
\altaffiltext{5}{UCLA, 8371 Mathematical Sciences, CA. 90095. USA. {\tt mcelwain@astro.ucla.edu}}
\altaffiltext{6}{Spitzer Fellow}

\label{firstpage}

\begin{abstract}
We present first results from a major program of methane filter photometry
for low-mass stars and brown dwarfs. The definition of a new methane filter
 photometric system is described. A recipe is provided
 for the differential calibration of methane imaging data
 using existing 2MASS photometry. 
 We show that these filters are effective in discriminating
 T dwarfs from other types of stars, and demonstrate this with
 Anglo-Australian Telescope observations using the IRIS2 imager.
 Methane imaging data and proper motions are
 presented for ten T dwarfs identified as part of the
 2MASS ``Wide Field T Dwarf Search'' -- seven of them initially identified
 as T dwarfs using methane imaging.
 We also present near-infrared moderate resolution 
 spectra for five T dwarfs, newly discovered by this technique. 
 Spectral types obtained from these spectra
 are compared to  those derived from both our methane filter observations, 
 and spectral types derived by other observers.
 Finally, we suggest a range of future programs to which
 these
filters are clearly well suited: the winnowing of T dwarf and Y dwarf
candidate objects coming from the next generation of near-infrared sky
surveys; the robust detection of candidate planetary-mass brown dwarfs in clusters;
the detection of T dwarf companions
to known L and T dwarfs via deep methane imaging; 
and the search for rotationally-modulated time-variable
surface features on cool brown dwarfs.

\end{abstract}

 
\section{Background}
\label{background}

 T dwarfs have the coldest photospheres (at least outside our Solar System)
 which are currently accessible to direct observation. With masses inferred
 to lie between $\sim 60-10$\Mjup, 
 they represent a class of object linking the properties of observable
 low-mass stars and brown dwarfs, with the properties
 of unobservable extrasolar planets. The defining feature of these objects is
 the presence of strong, broad methane absorptions in the near-infrared  at
 1.3-1.4, 1.6-1.8 and 2.2-2.5$\mu$m. 
 When first seen in the prototype T dwarf Gl\,229B \citep{nak1995}, 
 the distinctiveness of these spectroscopic features immediately
 suggested the need for a new spectral class. This ``T'' class was initially 
 explored in two separate spectral typing schemes by \citet{bu2002a} and 
 \citet{ge2002}, which are in the process of being unified in to a single
 ``hybrid'' system \citep{bu2003a}.
 
  The methane features that define T dwarfs are so broad and distinctive, that the use of
 dedicated filters to detect them was a logical next step. 
 Proof of concept observations of Gl\,229B \citep{ro1996} soon confirmed this expectation. 
  Many more
 T dwarfs have been discovered in the years since, and there has been
 some additional exploratory work on the use of methane filters \citep{he1999,ma2003} using this
  larger sample of T dwarfs. \citet{ma2004}, in particular, have defined a four filter
 narrow-bandpass photometric system in the atmospheric H window (1.45-1.8\,$\mu$m) designed
 for the detection and characterisation of young brown dwarfs in heavily reddened clusters.
 \citet{gol2004} and \citet{krist1998} have also presented results for methane imaging
 of nearby stars in a search for cool companions, based on a synthetic spectrum calibration.
 However,  near-infrared J,H,K imaging and near-infrared
 spectroscopy remain the mainstay of 
 T dwarf observational study. There has, as yet,
 been no large scale characterisation of observed T dwarfs using
 broad methane filters. Nor have methane filters yet
  been utilised where they are particularly powerful -- as a quick and efficient way of
 characterising  candidates arising from
 large photometric surveys as  ``T or non-T'', without the need for
 either photometric conditions or an infrared spectrograph.

 We therefore set out to more fully explore the use of methane filters 
 in the study of T dwarfs. Three steps were critical in this process:
 (a) defining a photometric calibration procedure for a 
 new photometric system (denoted \Ms, \Ml, and \Msl) using a set of broad
 methane filters with IRIS2  on the 3.9m Anglo-Australian
 Telescope (AAT); 
 (b) exploring the use of Two Micron All Sky Survey (2MASS; \citealt{sk1997,cu2003}) JHKs photometry 
   to calibrate methane filter  
 images differentially onto this \Ms, \Ml, \Msl\ photometric system,
 and (c) using the resulting differential methane imaging to search for T dwarfs
 amongst the lists of candidates being identified in the 2MASS Wide-Field
 T Dwarf Search (WFTS) -- a long-term program
 which has been conducting a search  over 74\% of the sky for T dwarfs
 in the 2MASS All-Sky database \citep{bu2003b,bu2003c,bu2004a,bu2004b}. 
 
 The selection criteria and selection
 procedures of the 2MASS WTFS are described in detail in \citet{bu2003b}. 
 The most challenging aspect of this program is the extreme {\em  rarity} of 
 T dwarfs in a shallow magnitude-limited
 survey.
 The first-pass photometric selection criteria produce over a quarter
 of a million potential T dwarfs from the 1.3 billion sources in the 2MASS Point
 Source Catalogue. But our expectation is that this
 sample will only contain $\sim20-30$ {\em actual} T dwarfs.
 Visual examination of catalogue images are carried out on all the first pass candidates
 to eliminate $\sim 99.5$\% of the faint background stars,
 proper-motion stars, closely separated visual binaries, 
  and general chaff that contaminate the sample.
 
 However, this still leaves $\sim 1000$ T dwarf candidates, which
 require further on-sky observations to winnow down to the much smaller
 number of actual T dwarfs expected. In the past, the procedure
 for doing this has been \citep{bu2002a}: near-infrared imaging to eliminate
 minor planets; followed by optical imaging to exclude background stars; followed
 by near-infrared
 spectroscopy to carry out final verification. The latter two steps are
 time consuming and expensive in telescope time, due to the faintness of these targets.
 Methane imaging offers
 the possibility of combining all three of these steps into one observation.
 A pair of 2-3 minute methane exposures
 on a 4m-class telescope can reveal; (a) whether there is still an infrared object
 present at the 2MASS position (i.e. eliminate uncatalogued minor planets); 
 (b) the colour of the object
 in the methane filters (which can determine its ``T or non-T'' nature and 
  estimate T spectral type); and (c) automatically provide
 proper motion data. Follow-up infrared spectroscopy, therefore,
 need only be carried out on a much smaller sample of high confidence
 T dwarf candidates, rather than the hundreds otherwise required.
 The greatly reduced data processing overhead of a pair of methane images, compared
 to infrared images, optical images and infrared spectroscopy, is also a
 significant advantage.

 This is Paper IV in the series arising from the WFTS. 
 Paper I \citep{bu2003b} describes the selection process
 for our T dwarf candidates in detail. Papers II-III \citep{bu2003c,bu2004a} 
 present initial T dwarf discoveries arising from the search.
 In this paper we describe the definition of a new \Ms, \Ml, \Msl\ 
 photometric system using methane filters, 
 provide sequences relating infrared JHK colours to
 these methane colours, and describe a procedure
 for using 2MASS data to differentially calibrate observations
 onto that system. 
 We present IRIS2 \Msl\ data and
 proper motions for new T dwarfs identified by this program -- five new T dwarfs,
 and a number of T dwarfs previously published by us \citep{bu2003c,bu2004a}, many of
 which were independently detected by our methane imaging in 2002-2003.
 In Section \ref{spectra} we present spectra for a number of these new T dwarfs, obtained with the
 IRIS2 instrument, and demonstrate that these spectra confirm our
 methane filter detections. Finally, Section \ref{discussion} discusses the extension
 of these techniques into other areas of T dwarf research.

\section{Methane Imaging}

\subsection{Imaging with IRIS2}
IRIS2 is an all-refracting 9 element focal-reducing 
collimator-camera, 
which is installed at the AAT's $f/8$ Cassegrain focus. It provides a 
final $f$-ratio at its HAWAII1 HgCdTe detector of $f/2.2$, or a 
plate scale of 0.4486\arcsec\ per pixel. This results in  a field-of-view 
7.7\arcmin\ on a side. The optical train is capable of delivering 1\,pixel 
full-width half-maximum images
over the whole field of view. However, as with all focal reduction systems 
of this type, it introduces astrometric distortion into
the detected images. IRIS2's astrometric distortion is such that the
plate scale in the corners of the detector is $\sim$1\% smaller than
that at the field centre. It is quite precisely represented as a radial
distortion, which can be parametrised by a quartic polynomial of the form

\begin{equation}
r = r^\prime(1 - 2.4988\times10^{-6}r^\prime -4.4466\times10^{-11}{r^\prime}^3)
\end{equation}

where $r^\prime$ is the radius in observed pixels from a central
pixel ($x_0$,$y_0$)=(516.86,515.02), and $r$ is the radius in pixels from ($x_0$,y$_0$)
in an ideal undistorted co-ordinate system with plate scale 
$0.4486\arcsec$ per pixel.\footnote{See the IRIS2 web pages at 
{\tt\url{http://www.aao.gov.au/iris2/iris2.html}} for
details.} 

\begin{figure}
   \centering\includegraphics[width=80mm]{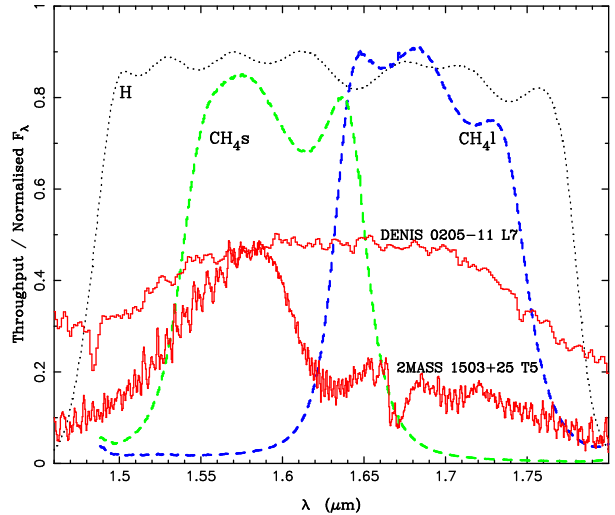}
   \caption{IRIS2 methane filters. The heavy dashed lines show the
            measured bandpasses of the IRIS2 \Ms\ and \Ml\ filters. The cut-on, central 
            and  cut-off wavelengths for the \Ms\ and \Ml\ filters are 1.520, 1.570, 
            1.620\,$\mu$m and 1.640, 1.690, 1.740\,$\mu$m (respectively). For comparison 
            Mauna Kea Observatories H filter bandpass (also installed in IRIS2)
            is plotted as a  dotted line. Spectra of the T5 dwarf 2MASS J15031961+2525196
            (lower trace: SPEX data from \citet{bu2004a}),
            and the L7 dwarf DENIS-P J0205.4-1159 (upper trace; CGS4 data from \citet{le2001}).
            The two spectra have been normalised at 1.57$\mu$m.}
   \label{methane1}
\end{figure}

IRIS2 contains a filter set acquired as part of one of the ``Mauna Kea
Observatories'' infrared filter consortia.
J,H,K and Ks filters were acquired from OCLI of Santa Rosa, CA in 1998,
and a set of narrow- and intermediate-band filters were acquired from NDC 
Infrared Engineering, Essex, UK between 2000 and 2002. The J,H,K,Ks filters
were manufactured to the ``Mauna Kea Observatories Near-Infrared Filter Set''
specifications of \cite{to2002}. The narrow- and intermediate-band
set are as specified by 
A. Tokunaga\footnote{\tt \url{http://www.ifa.hawaii.edu/\%7Etokunaga/NB\_special\_ordersorting.html}}.
In particular, the latter set
includes a pair of methane filters, which we denote \Ms\ and \Ml, which are available in many
other infrared instruments at telescopes around the world. 
Figure \ref{methane1} shows the band-passes for these filters, superimposed
on  T and L dwarf spectra. The \Ml\ filter samples the
the broad 1.6-1.8$\mu$m methane absorption bands seen in T dwarfs, while the
\Ms\ filter samples pseudo-continuum outside the methane band -- though as
for all cool dwarf photospheres this pseudo-continuum is itself the result
of significant molecular absorption -- in this case largely H$_2$O.

\subsection{IRIS2 Observations}

Imaging observations in methane filters were acquired (in 
combination with a variety of other observing programs)
on 29 nights in the period 2002 March 31 to 2004 March 9 with IRIS2. Observing
conditions on these nights were highly variable -- from good seeing
in photometric conditions, to poor seeing in partial cloud.
(In many cases
methane observations were carried out as a ``poor conditions'' backup 
to other observing  programs.)
Objects targeted for observation over the course of this program include
known T dwarfs, known L dwarfs, a large number of candidate T dwarfs
deriving from the 2MASS WFTS, and (in
photometric conditions) UKIRT ``Faint Standard'' (FS) stars \citep{ha2001}.

Observations with these filters are carried out by obtaining a series
of dithered individual exposures in each filter. 
All these observations are performed under the control of an IRIS2 observing
sequence, providing dithers in either fixed or pseudo-random patterns.
As data is taken it is processed on-line using an
implementation of the ORAC-DR data reduction 
pipeline\footnote{\tt \url{http://www.oracdr.org/}}. 
ORAC-DR is a generic data reduction pipeline originally created at the Joint Astronomy Centre, Hawaii, 
for use with various UKIRT and JCMT instruments. 
It collects sets of observations of a target in a single filter into a
{\em group}. It generates a first pass flat-field for the {\em group}
by suitably normalising and creating a median image. It uses these first-pass flattened 
data to detect objects, mask them out, and flag bad pixels. It then creates
a final flat field by taking the median of this normalised and masked data, and
applies it to the raw data to create a {\em flattened group}.
The flattened group images are re-sampled to remove astrometric distortion,
objects are re-detected and offsets between images estimated using these
detections. The images are then re-sampled again onto a uniform co-ordinate
system to create a single {\em final} image. All of this processing is
carried out on a dedicated data-reduction 3.3Ghz Linux PC, resulting in
fully processed images within 30s of an observing sequence completing -- usually
before the telescope has even been moved to the next observing target.

%
%

\subsection{A Methane Photometric System}
\label{absolute}

To calibrate the methane colours of T dwarfs, extensive observations of both
UKIRT FS stars, and known M, L and T dwarfs were obtained on
at least three photometric nights over the course of our observing program: 
2002 October 21; 2003 January 25; and 2004 March 1. 
There are no pre-existing photometric standards for use with such filters, 
so we elected to define our own photometric system based on the
Mauna Kea Observatories (MKO) H-band photometry of 
UKIRT FS stars\footnote{\label{MKOFS}\tt \url{http://www.jach.hawaii.edu/UKIRT/astronomy/calib/phot\_cal/}}.
Additional JHK photometry on the same MKO system for a number of M, L and T dwarfs
was obtained from \citet{le2002} and \citet{kn2004}. Where MKO JHK photometry
for M, L and T dwarfs observed in the methane filters was not available,
we converted 2MASS All Sky photometry \citep{cu2003} to the MKO system using
published spectral types, and the conversion functions of \citet{st2004}.

Aperture photometry was measured for all the targets observed on these nights.
For bright targets, large photometric apertures (10-15\arcsec\ in radius)
were chosen, resulting in $<$0.005 magnitudes in flux being missed
outside the aperture. For fainter targets, photometry was performed in
smaller (2-4\arcsec\ radius apertures), and bright objects in the same image
were used to determine aperture corrections to the same large apertures as used
for brighter objects.
Uncertainties due to the standard-error-in-the-mean of these aperture corrections 
(estimated on an image-by-image basis) and the $<$0.005\,mag. flux potentially missed 
outside the 
outer aperture, were carried through all subsequent calculations.

We adopted UKIRT FS stars of spectral types  
A, F and G as fundamental standards. This spectral type range was chosen because (as we show below)
it shows essentially zero \Ms$-$\Ml\ colour variation.
We assign to these UKIRT FS stars magnitudes in the \Ms\ and \Ml\
filters identical to their MKO H magnitudes, and from this determine
photometric calibrations (both airmass corrections and zero-points) on each night.
These are reported in Table \ref{photomcalib}. In all but one case, the \Ms\ and \Ml\ 
airmass corrections and zero-points were consistent with being the same (on a given
night), so average values were adopted and are reported in the
table. The exception was 2004 March 1, where the airmass corrections
were not consistent with a single value, and so different values were used.
Both airmass corrections on this night were also significantly larger than those
obtained on the other two nights, and the photometric zero-point has significantly
larger uncertainties. This night was warm, humid and somewhat hazy, which is thought
to be the reason for the larger extinction and poorer  photometric accuracy.
There is also significant variation from night-to-night in the photometric zero-point.
These variations most likely reflect the fact these filters sample the edges of
the atmospheric water vapour absorption bands which define the
H window (at least at the Siding Spring 
site). Night-to-night variations in the amount of water vapour absorption will
result in differing zero-points. However, since the critical parameter these
filters measure is the {\em difference} between \Ms\ and \Ml, so long as observations
in the two filters are closely spaced in time and airmass, these variations are
essentially irrelevant.

\begin{table}
\begin{center}
  \caption{\Ms\ and \Ml\ photometric parameters\label{photomcalib}}
  \begin{tabular}{@{}lcccc@{}}
  \colrule\colrule
   Night        &\multicolumn{2}{c}{Airmass Correction, $A$\tablenotemark{a}}
                                                 &\multicolumn{2}{c}{Zeropoint, $ZP$} \\
                &    \Ms\       & \Ml            & \Ms\ & \Ml \\
  \colrule
2002 October 21 & 0.03$\pm$0.03 & 0.03$\pm$0.03  & 10.46$\pm$0.03 & 10.46$\pm$0.02\\
2003 January 25 & 0.03$\pm$0.03 & 0.03$\pm$0.03  & 10.76$\pm$0.02 & 10.76$\pm$0.02\\
2004 March 1    & 0.42$\pm$0.01 & 0.45$\pm$0.01  & 10.35$\pm$0.06 & 10.35$\pm$0.06\\
  \colrule
\end{tabular}
\tablenotetext{a}{ Calibrated magnitude = $-2.5{\log}_{10}(counts/t_{exp})-A\,{\times}\,airmass + 31 - ZP$.}
\end{center}
\end{table}

\subsection{Methane Photometry}

\subsubsection{\Msl\ and Spectral Type}
\label{methanespectype}

\begin{figure*}
   \centering\includegraphics[angle=-90,width=140mm]{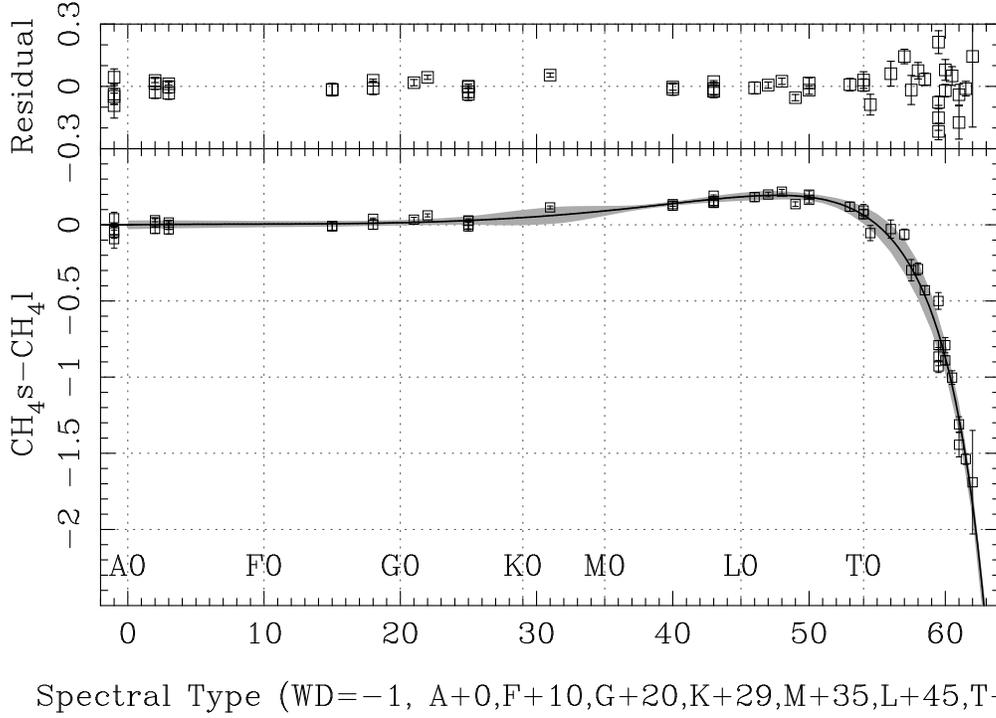}
   \caption{IRIS2 methane colour (\Ms$-$\Ml) as a function of A-T spectral type.
            The uncertainties plotted are the combination of photon-counting
            uncertainties, aperture correction uncertainties and photometric
            calibration uncertainties. Typical uncertainties on spectral types
            (not plotted) are $\pm$0.5. See the text for references to the spectral
            type sources and for the form of the plotted parametrisation.
            Root-mean-square (rms) scatter about the parametrisation (grey shading) is 0.07\,mag 
            for the whole range, 0.04\,mag for L0-T2 dwarfs and 0.11\,mag for
            T dwarfs.}
   \label{meth_spt}
\end{figure*}

These observations therefore provide photometry 
in \Ms, \Ml\ and, most importantly, \Msl\
colour. Figure \ref{meth_spt} shows \Msl\ colour as a function of
spectral type. Spectral type is parametrised numerically as $n$, such 
that for A to T type stars $n$ is the 
spectral subtype plus a constant : 0 for an A dwarf; 10 for F; 20 for G;
29 for K; 35 for M; 45 for L and 54 for T. 
While it was not our intention in this study to study the detailed methane
spectroscopy of white dwarfs, nonetheless three UKIRT FS 
white dwarfs were observed during the course of this program, which we have put 
at the ``placeholder'' position of $n=-1$.
The objects plotted (and tabulated in Table \ref{methdata}) are: UKIRT FS stars with spectral types A-K stars from \citet{ha2001}
and for white dwarfs from \citet{ms1999}; M dwarfs with \citet{le2002}
spectral types; L dwarfs with \citet{ki2000} types, except for
2M1045-0149 \citep{gizis2002} and 
DEN1539-0520, which is based on a Keck 
optical spectrum analysed 
so as to place it on the \citet{ki2000} system (D. Kirkpatrick, private communication); 
and T dwarfs with spectral types on a variant of the hybrid T typing scheme 
which was preliminarily proposed by \citet{bu2003a}, but modified
to use slightly different indices and spectrscopic reference standards
as described in Section \ref{spectral_classification}.
Table \ref{methdata} also provides the average \Msl\ colours for all the objects in
Figure \ref{meth_spt}, along with the spectral types  and JHK colours adopted, so that
others using such filters can calibrate their data onto a uniform system. 

As suggested above, the variation in \Msl\ colour across the
A, F and G spectral types is negligibly small, which is why these stars
were chosen to define the photometric system.  K, M and L dwarfs
have slightly positive \Msl\ colours (i.e. they are red in these filters), 
while T dwarfs show a marked trend to become bluer with increasing
T spectral type. Figure \ref{meth_spt} also shows a parametrisation for this relationship, as
follows,

\begin{eqnarray}
\mathrm{CH}_\mathrm{4}\mathrm{s}-\mathrm{CH}_\mathrm{4}\mathrm{l} &=& 
n(0.0087 + 6.176{\times}10^{-6}\,n^2 \nonumber\\
&&+ 1.202{\times}10^{-9}\,n^4 - {{0.519}\over{68.5-n}})
\label{methane_to_spectraltype}
\end{eqnarray}.

While this function is numerically valid to $n<68.5$,
it is only constrained by data in the range $0<n<62$ (i.e.. the white dwarfs do not
constrain the fit). Root-mean-square (rms)
scatter about this fit is
0.07\,mag over the whole spectral type range. More critically
the rms scatter is 0.11\,mag for T dwarfs and 0.04\,mag for L0-T2 dwarfs.

\subsubsection{\Msl\ and JHK colours}
\label{colours}

While the relationship between \Msl\ colour is 
useful for assigning T spectral types, it is not the only
tool for examining the potential T dwarfs in large photometric surveys
like 2MASS or the forthcoming
UKIRT Infrared Deep Sky Survey (UKIDSS\footnote{\tt \url{http://www.ukidss.org/}} \citet{ham03} ). 
These surveys will also produce JHK photometry,
providing additional data on whether or not a given object is likely
to be a T dwarf.

\begin{figure*}
   \centering\includegraphics[width=90mm,angle=270]{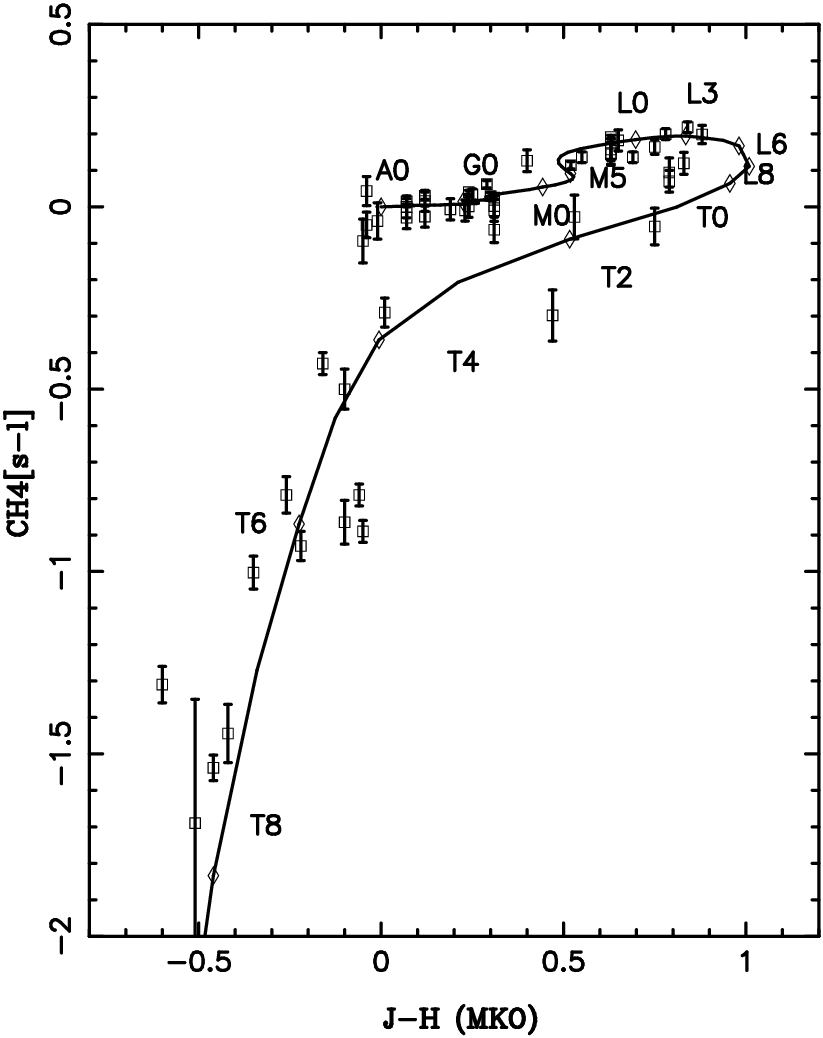}
   \centering\includegraphics[width=90mm,angle=270]{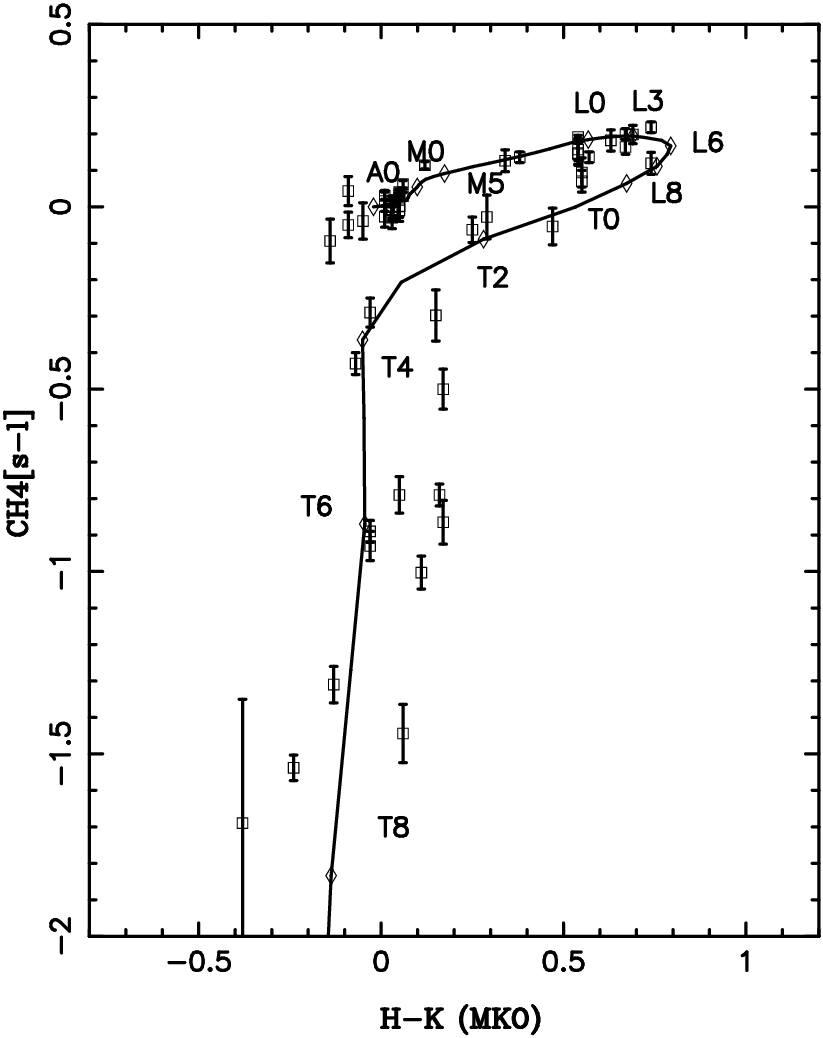} \\[0.5cm]
   \centering\includegraphics[width=90mm,angle=270]{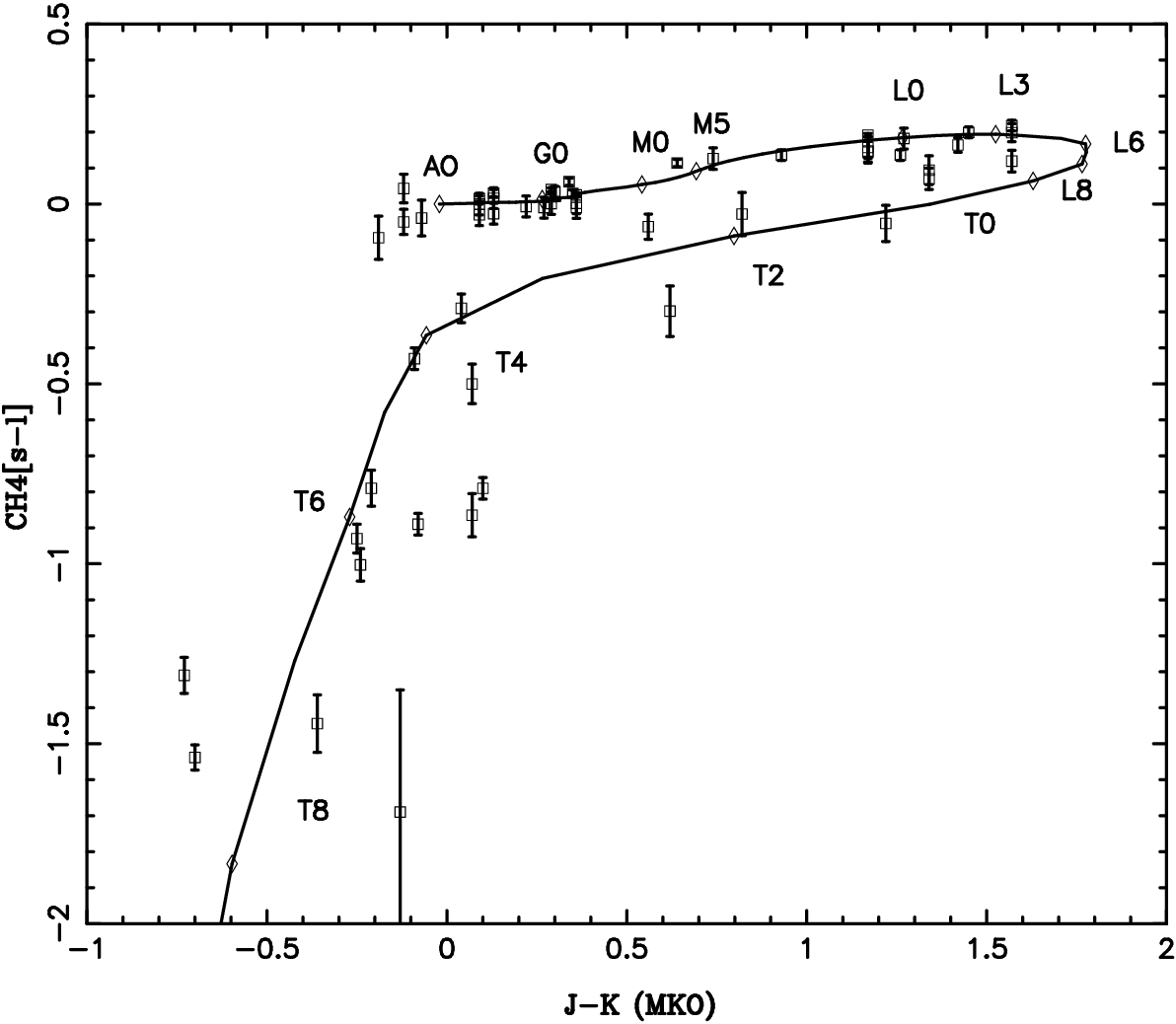}
   \caption{IRIS2 methane colour (\Ms$-$\Ml) as a function of J-H, H-K, and J-K (MKO) colour.
            The \Msl\ uncertainties plotted are the combination of photon-counting
            uncertainties, aperture correction uncertainties and photometric
            calibration uncertainties. JHK photometry and uncertainties are
            from UKIRT WWW pages, MKO photometry from \citet{le2002} or
            2MASS photometry converted to the MKO system using the conversion
            relations of \citet{st2004}. 
            The overplotted tracks and spectral types ({\em solid lines and open
            diamonds}) are produced from the \Msl\ versus
            spectral type relation of Fig. \ref{meth_spt}, and the JHK-colour versus spectral type
            relations of Fig. \ref{jhk_spt} and Table \ref{meth_jhk_table}. Note the four white dwarf observations
            which when plotted in this plane, lie just to the left of the A0 end of the plotted
            sequence.
            }
   \label{meth_jhk}
\end{figure*}

Figure \ref{meth_jhk} shows plots of \Msl\ colour as a function
of J-H, H-K and J-K colour (on the MKO photometric system) for the
objects (except the white dwarfs) presented in Figure \ref{meth_spt}. The JHK photometry
for these objects comes from (in order of preference) : the UKIRT MKO
photometric standards 
database$^{\ref{MKOFS}}$;
from \citet{le2002}; from \citet{kn2004}; or by converting 2MASS All Sky data
into the MKO system using the conversion relations of \citet{st2004}.
Clearly the dwarf sequence has a complex behaviour in these diagrams, with the
added complication of significant scatter among the T dwarfs. 

To try and clarify this behaviour we have combined the \Msl versus spectral type
relations derived above, with JHK versus spectral type sequences for dwarf
stars. The latter have been derived using the spectral types of \citet{ha2001}
for UKIRT MKO photometric standards, supplemented by published
M, L and T types (see Sections  \ref{methanespectype} and 
\ref{spectral_classification} for details and references)
for a number of cool dwarfs 
with existing MKO JHK photometry \citep{le2002,kn2004}. The resulting
sequences are shown in Figure \ref{jhk_spt}. Tests with polynomials of both low
and high order show that these curves are not readily amenable to simple parametrisation,
so we have parametrised the J-K and H-K sequences by binning the available data in
spectral type (large open squares in the figure), and interpolating
a cubic-spline through these data points (solid line in the figure). 

These near-infrared sequences, combined with the \Msl\ versus spectral type
sequences derived above, produce ``tracks'' in the \Msl\ versus colour planes, 
which are
shown in Figure \ref{meth_jhk}. These tracks tell us (among other things)
that to discriminate between early T dwarfs and the M-L dwarfs, 
\Msl\ photometry must be sufficiently
precise to reveal the 0.2\,mag gap between these two classes of object.
The sequences are tabulated in Table \ref{meth_jhk_table}.

One notable feature of these tracks is the inflection at J-H$\approx$0.5 in
the J-H versus \Msl\ plane, which is simply the mapping into J-H versus \Msl, 
of the well known inflection in the J-H versus H-K diagram, which is produced 
as giants bifurcate from dwarfs at mid-M spectral types (see for example \citet{bb1988}, 
Figure A3).

\begin{figure}
%
%
   \centering\includegraphics[width=70mm]{f5.eps}
   \caption{\Msl\ dwarf sequence as a function of J-H 
            in the MKO and 2MASS systems.
            MKO curve ({\em dot-dashed line}) comes from the parametrisations
            in Fig. \ref{meth_jhk} and 2MASS ({\em dashed line})
            curve is the MKO curve transformed using the conversions
            in the 2MASS Explanatory
            Supplement and \citet{st2004}. The solid lines are
            polynomial fits to ranges (J-H)$_\mathrm{MKO}=-$0.05-0.5 and
            0.6-1.1 which are non-degenerate and useful for determining the 
            \Msl\ zero point of observations with 2MASS photometry.}
   \label{JH_MKO_2MASS}
\end{figure}

\begin{figure*}
   \centering\includegraphics[width=100mm]{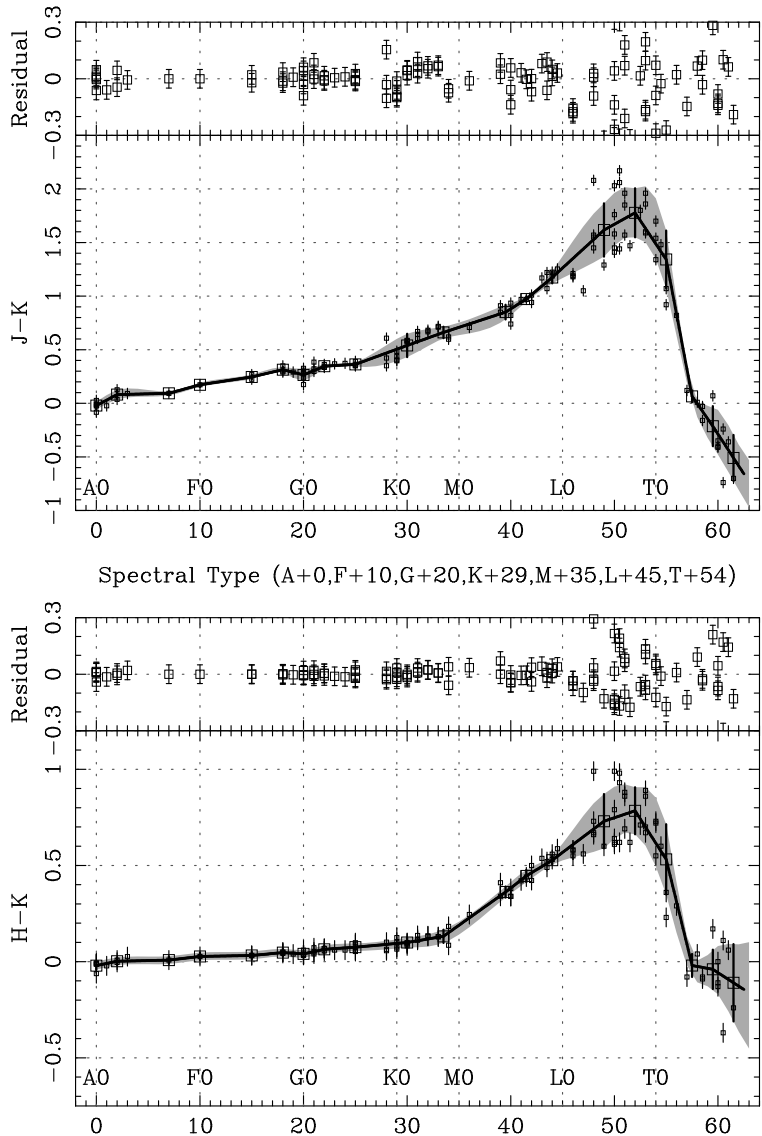}
   \caption{J-K and H-K (MKO) as a function of A-T spectral type.
            JHK photometry and uncertainties are
            from UKIRT WWW pages, and MKO photometry from \citet{le2002} for M, L and 
            T dwarfs ({\em small squares}). These data have been binned by spectral type
            ({\em large squares}), which have  themselves been fitted with
            a cubic-spline interpolating function ({\em solid line}). 
            The shaded region shows the 1-$\sigma$ rms scatter about the binning. 
            These tracks are summarised in Table \ref{meth_jhk_table}.
            }
   \label{jhk_spt}
\end{figure*}

\begin{table*}
  \begin{center}
  \caption{Spectral type, Methane and JHK colour sequences\label{meth_jhk_table}}
  \begin{tabular}{@{}lccccccc@{}}
  \colrule\colrule
    &      &\multicolumn{3}{c}{MKO} & \multicolumn{3}{c}{2MASS} \\
SpT &\Msl\tablenotemark{a}  
             & J-H\tablenotemark{b}
                       &H-K\tablenotemark{b}
                                 &J-K\tablenotemark{b}
                                           &J-H\tablenotemark{c}   
                                                     &H-K\tablenotemark{c}   
                                                               &J-K\tablenotemark{c} \\ 
  \colrule
A0 &$+$0.000 &$+$0.000 &$-$0.021 &$-$0.021 &$-$0.038 &$-$0.002 &$-$0.023 \\ 
A5 &$+$0.004 &$+$0.087 &$+$0.006 &$+$0.093 &$+$0.063 &$+$0.021 &$+$0.095 \\ 
F0 &$+$0.005 &$+$0.146 &$+$0.027 &$+$0.173 &$+$0.131 &$+$0.039 &$+$0.178 \\ 
F5 &$+$0.007 &$+$0.214 &$+$0.032 &$+$0.246 &$+$0.209 &$+$0.044 &$+$0.254 \\ 
G0 &$+$0.014 &$+$0.225 &$+$0.040 &$+$0.265 &$+$0.222 &$+$0.051 &$+$0.274 \\ 
G5 &$+$0.028 &$+$0.288 &$+$0.075 &$+$0.363 &$+$0.295 &$+$0.081 &$+$0.375 \\ 
K0 &$+$0.047 &$+$0.407 &$+$0.092 &$+$0.499 &$+$0.432 &$+$0.096 &$+$0.516 \\ 
K3 &$+$0.067 &$+$0.504 &$+$0.110 &$+$0.614 &$+$0.545 &$+$0.112 &$+$0.636 \\ 
M0 &$+$0.091 &$+$0.519 &$+$0.174 &$+$0.693 &$+$0.562 &$+$0.167 &$+$0.718 \\ 
M2 &$+$0.110 &$+$0.495 &$+$0.252 &$+$0.747 &$+$0.534 &$+$0.235 &$+$0.774 \\ 
M4 &$+$0.129 &$+$0.486 &$+$0.340 &$+$0.826 &$+$0.524 &$+$0.311 &$+$0.856 \\ 
M6 &$+$0.149 &$+$0.512 &$+$0.426 &$+$0.938 &$+$0.554 &$+$0.386 &$+$0.972 \\ 
M8 &$+$0.168 &$+$0.595 &$+$0.495 &$+$1.090 &$+$0.650 &$+$0.446 &$+$1.129 \\ 
L0 &$+$0.184 &$+$0.698 &$+$0.568 &$+$1.266 &$+$0.853 &$+$0.538 &$+$1.391 \\ 
L2 &$+$0.193 &$+$0.792 &$+$0.651 &$+$1.443 &$+$0.958 &$+$0.579 &$+$1.537 \\ 
L4 &$+$0.190 &$+$0.878 &$+$0.723 &$+$1.601 &$+$1.056 &$+$0.633 &$+$1.689 \\ 
L6 &$+$0.167 &$+$0.981 &$+$0.794 &$+$1.775 &$+$1.173 &$+$0.706 &$+$1.879 \\ 
L8 &$+$0.111 &$+$1.009 &$+$0.756 &$+$1.765 &$+$1.218 &$+$0.685 &$+$1.903 \\ 
T0 &$+$0.064 &$+$0.956 &$+$0.673 &$+$1.629 &$+$1.176 &$+$0.615 &$+$1.791 \\ 
T1 &$-$0.001 &$+$0.810 &$+$0.532 &$+$1.342 &$+$1.041 &$+$0.490 &$+$1.531 \\ 
T2 &$-$0.089 &$+$0.517 &$+$0.281 &$+$0.798 &$+$0.761 &$+$0.257 &$+$1.018 \\ 
T3 &$-$0.207 &$+$0.211 &$+$0.055 &$+$0.266 &$+$0.470 &$+$0.050 &$+$0.520 \\ 
T4 &$-$0.365 &$-$0.006 &$-$0.051 &$-$0.057 &$+$0.269 &$-$0.036 &$+$0.233 \\ 
T5 &$-$0.579 &$-$0.126 &$-$0.047 &$-$0.173 &$+$0.167 &$-$0.012 &$+$0.156 \\ 
T6 &$-$0.870 &$-$0.225 &$-$0.045 &$-$0.270 &$+$0.088 &$+$0.011 &$+$0.099 \\ 
T7 &$-$1.270 &$-$0.340 &$-$0.083 &$-$0.423 &$-$0.004 &$-$0.007 &$-$0.012 \\ 
T8 &$-$1.834 &$-$0.460 &$-$0.137 &$-$0.597 &$-$0.100 &$-$0.042 &$-$0.142 \\ 
  \colrule
  \end{tabular}
\end{center}
\tablenotetext{a}{Sequence represented by equation \ref{methane_to_spectraltype}.}
\tablenotetext{b}{Spline sequences as shown in Fig \ref{jhk_spt}.}
\tablenotetext{c}{MKO sequences converted to 2MASS system as described in the text.}
\end{table*}

\subsection{Differential Methane Photometry}
\label{differential}

While photometry is essential to underpin our determination
of the
relationship between near-infrared colour, spectral type and \Msl\ colour,
the observational overheads in obtaining such data are large,
leading us to ask ``Can we make do with differential
photometry?''. Since both our bands can be observed
nearly simultaneously,
at identical airmasses in similar seeing
and transparency conditions, systematic uncertainties
in aperture corrections due to seeing variations, extinction corrections
due to airmass variations, and (to some extent) zero-point
corrections due to transparency variations, can all be cancelled out.
The availability of the 2MASS All Sky 
data provides an additional incentive to the differential approach, since
H-band calibration photometry is available over
almost 100\% of the sky.

Our \Msl\ photometric system is defined based on MKO H photometry
for standard stars of spectral type A, F and G. Or equivalently in
colour space -0.021$<$(J-K)$_\mathrm{MKO}$$<$0.456, 
-0.021$<$(H-K)$_\mathrm{MKO}$$<$0.086, 
or 0.00$<$(J-H)$_\mathrm{MKO}$$<$0.370. To use 2MASS photometry
to calibrate the same data, we can transform these colour ranges into
the appropriate 2MASS ranges using the conversions provided by  the 
2MASS All-Sky Survey Explanatory 
Supplement\footnote{\tt\url{http://www.ipac.caltech.edu/2mass/releases/allsky/doc/sec6\_4b.html}} for
A-M dwarfs, supplemented by the conversions of \citet{st2004} for
L-T dwarfs. Figure \ref{JH_MKO_2MASS} shows the \Msl\ versus 
(J-H)$_\mathrm{MKO}$ relation derived above, along with the comparable \Msl\ versus 
(J-H)$_\mathrm{2MASS}$ relation obtained by transforming MKO colours to
2MASS colours. (J-H has been chosen for this plot as it is the
colour most commonly available for objects in the 2MASS All Sky
catalogue -- similar relations have been derived for the other JHK colours). 
In principle, therefore, one could select
background stars in a pair of \Ms,\Ml\ frames with 2MASS colours
corresponding to A-G spectral types -- $-$0.023$<$(J-K)$_\mathrm{2MASS}$$<$0.472, 
$-$0.002$<$(H-K)$_\mathrm{2MASS}$$<$0.091, 
or $-$0.038$<$(J-H)$_\mathrm{2MASS}$$<$0.390.
Unfortunately, even with a field of view as large as that of
IRIS2 (7.7\arcmin\ on a side), this often restricts the number of
objects available for calculating a zero point to less than 2
per field.

However, Fig. \ref{JH_MKO_2MASS} indicates that
a much larger range of colours could be used to calculate
a zero-point, if one was able to parametrise the variation
in \Msl\ as a function of 2MASS colour. The solid
lines in Figure \ref{JH_MKO_2MASS} represent a parametrisation for (J-H)$_\mathrm{2MASS}$
which has been adopted to do this. We avoid the region 0.5$<$(J-H)$_\mathrm{2MASS}$$<$0.6
where the sequence is degenerate, and fit two separate quadratics
to the regions -0.05$<$(J-H)$_\mathrm{2MASS}$$<$0.5,
\begin{eqnarray}
\mathrm{CH}_\mathrm{4}\mathrm{s}-\mathrm{CH}_\mathrm{4}\mathrm{l} &=& 
 -0.0015 - 0.0.00864\,(\mathrm{J-H}) \nonumber\\
 && + 0.2569263\,(\mathrm{J-H})^2
\end{eqnarray}
and 0.6$<$(J-H)$_\mathrm{2MASS}$$<$1.1,
\begin{eqnarray}
\mathrm{CH}_\mathrm{4}\mathrm{s}-\mathrm{CH}_\mathrm{4}\mathrm{l} &=& 
-0.0378 + 0.47851\,(\mathrm{J-H}) \nonumber\\
&& - 0.249901\,(\mathrm{J-H})^2.
\end{eqnarray}

Of course, these relations are also
degenerate with the T dwarf branch of the sequence.
We therefore adopt a two-stage calibration approach --
by rejecting from the second pass
any objects within 10\arcsec\ of any known or suspected T dwarfs 
in the field, and any 
objects with \Msl\ $<-0.2$, we eliminate any T dwarfs which might
contaminate this calibration. 

\begin{figure}
   \centering\includegraphics[width=60mm]{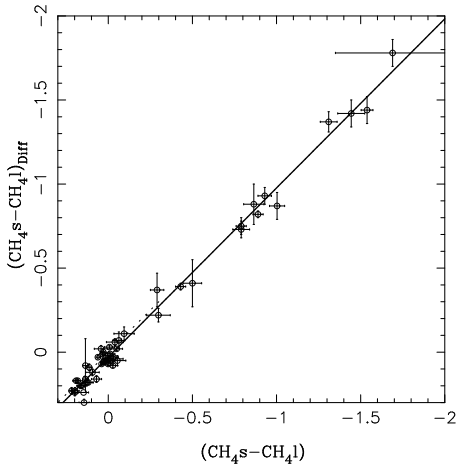}
   \caption{(\Msl) measurements plotted against 
            (\Msl)$_{Diff}$ measurements obtained from the same data frames.
            The dotted line is the one-to-one relationship, while the solid line
            is the least squares fit described in the text. The scatter in 
            (\Msl)$_{Diff}$ about the fit is 0.05\,mag.}
      \label{abs_diff}
\end{figure}

The following procedure, therefore, enables the
use of the ensemble of background stars in a field-of-view to derive
a robust differential \Msl\ in even non-photometric conditions : 
\begin{enumerate}
\item Select and photometer stars in \Ms\ and \Ml\ images. We carried out this
      step using the Starlink implementation of
      Sextractor\footnote{\tt http://terapix.iap.fr/soft/sextractor/index.html\label{sextractor}}
      \citep{Bertin96}.
      It is important to note that
      the detailed parameters used in this process are not critical, since
      the only aim is to obtain robust {\em relative} photometry for all
      the objects in a field of view.
\item Obtain 2MASS JHK photometry in the field of view -- these can be downloaded
      from the GATOR 
      server at IRSA\footnote{\tt http://irsa.ipac.caltech.edu/applications/Gator/}.
\item Match the \Ms, \Ml\ and 2MASS objects. Our implementation of this step
      was carried out using Michael Richmond's 
      {\tt match} code\footnote{\tt http://spiff.rit.edu/match/\label{match}}.
\item Avoiding any known or suspected T dwarfs
      within the field of view, select matched objects with -0.05$<$(J-H)$_\mathrm{2MASS}$$<$0.5 and 
      0.6$<$(J-H)$_\mathrm{2MASS}$$<$1.1 and uncertainty in 2MASS magnitude less than
      0.1\,mag. Then use equations 2 and 3 to estimate the \Msl\ for
      these objects based on their J-H colour, and 
      derive a first pass zero-point \Msl\ calibration.
\item Repeat this step, further rejecting any objects with \Msl\ $<-0.2$ from the 
      first pass.
\item Apply this zero-point calibration to place data on the \Msl\
      system.
\end{enumerate}
We have found that a typical IRIS2 field of view
will contain between 5 and
15 suitable background objects, leading to zero point calibrations with a
standard error in the mean of $<$0.05\,mag.

Figure \ref{abs_diff} shows a plot of the photometry (used to define
our methane photometric system in Section \ref{absolute}) versus differential methane
photometry calculated for the same observations using the procedure above. The dotted line
shows the one-to-one relationship between the two sets of measurements, while the
solid line is a least squares fit between the two, with the following form
\begin{eqnarray}
(\mathrm{CH}_\mathrm{4}\mathrm{s}-\mathrm{CH}_\mathrm{4}\mathrm{l})_{Diff} &=&
   0.029\pm0.001 \nonumber\\
   &&+ (1.006\pm0.004) (\mathrm{CH}_\mathrm{4}\mathrm{s}-\mathrm{CH}_\mathrm{4}\mathrm{l})
\end{eqnarray}
The rms scatter in differential \Msl\ about this fit is 0.05\,mag.
These results indicate that differential \Msl\
photometry can be used with confidence at the 0.05\,mag level.

\subsection{Methane Imaging Results}
\label{methaneresults}

With a typical methane imaging observation  consisting of 3-5
dithered observations in each of the methane filters, and
exposure times of 30-60\,s, a complete WFTS candidate sequence 
takes $\sim$5 minutes to complete. 
The ORAC-DR system is able to process these data to publication standard
on the fly, and a perl script automates the processing described in the previous
sections, allowing nearly on-line analysis of each observation during each
night (or at the very latest, on the following afternoon).
This has the advantage of allowing almost immediate IRIS2 spectroscopic
follow-up of T dwarfs identified by our methane imaging.

To date, we have observed 
508 candidate T dwarfs identified
as part of the WFTS -- of these 61 revealed no
infrared counterpart to the original 2MASS detection, while the remainder
were detected in the \Ms\ and \Ml\ bands. Of these detected objects:
10 have \Msl\ colours indicating they are T dwarfs, and have been
spectroscopically confirmed as T dwarfs;  2 have colours suggesting they may
be T dwarfs, and have been ruled out by spectroscopy; and 3 objects have \Msl\ 
colours suggesting they may be early T dwarfs and are awaiting
spectroscopy to confirm their status. Of the 10 T dwarfs detected
by methane imaging from the WFTS,  seven (2M0034-33, 2M1122-35,
2M1114-26, 2M0939-24, 2M0949-15, 2M1828-48, 2M2331-47) were first confirmed
as T dwarfs by methane imaging, while a further three were imaged
in methane filters after their spectra had been acquired elsewhere
\citep{bu2003c,bu2004a}. Finding charts for all of these T dwarfs are
shown in Figure \ref{charts}.

\begin{figure*}
   \centering\includegraphics[width=90mm]{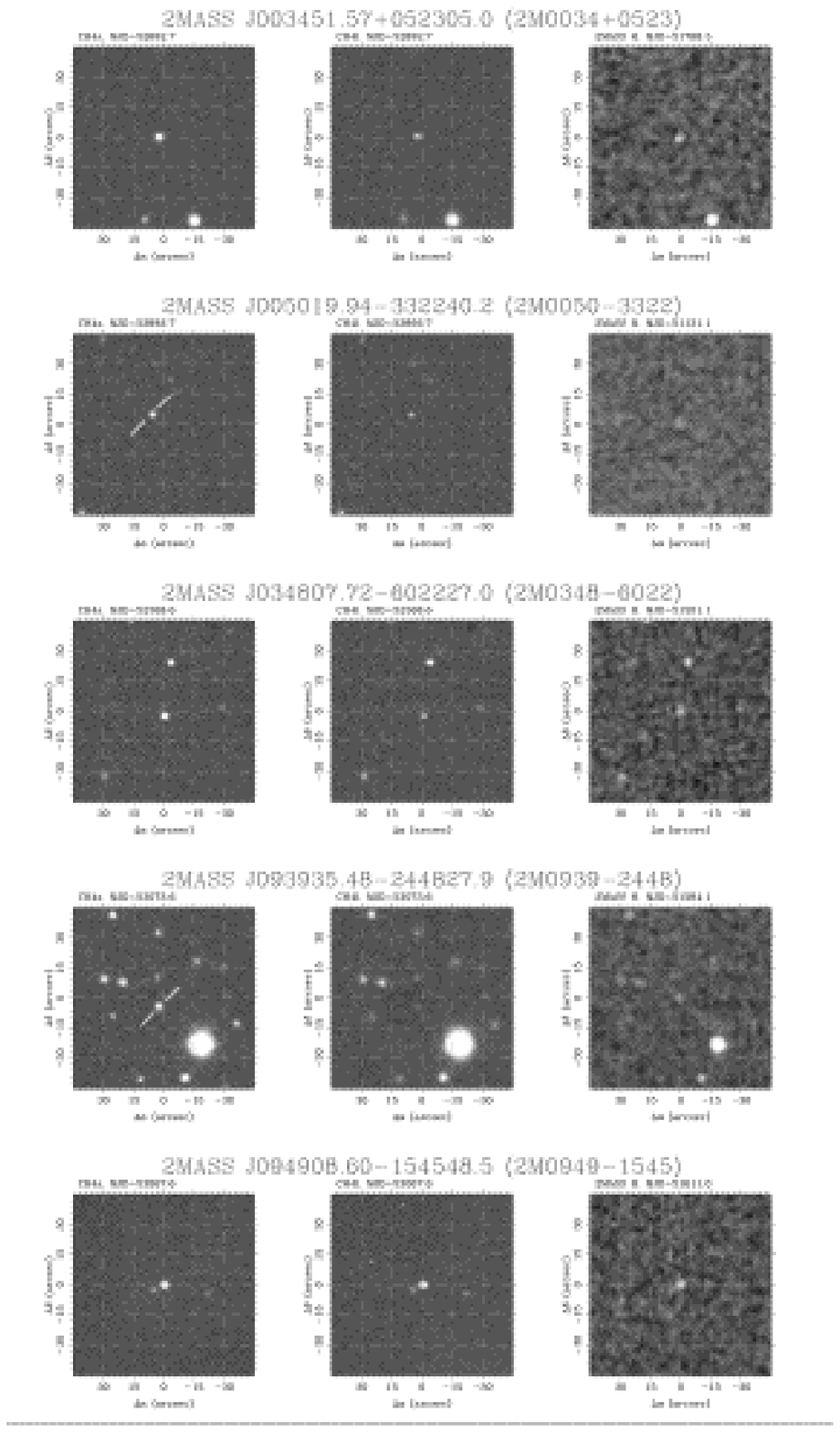}
   \caption{Finding charts for WFTS objects listed in Table \ref{diff_table}. For
            each object, images 3\arcmin\
            on a side, centred on the 2MASS Point Source Catalog (PSC) position
            indicated, in \Ms, \Ml\ and 2MASS H are shown. Epochs are indicated by
            their Modified Julian Dates. Also shown superimposed on the H-band 
            charts are all
            the 2MASS PSC sources in each field of view. The WFTS T dwarfs are highlighted
            in the methane images where proper motions make identification
            unclear.}
      \label{charts}
\end{figure*}
\begin{figure*}
   \figurenum{7}
   \centering\includegraphics[width=90mm]{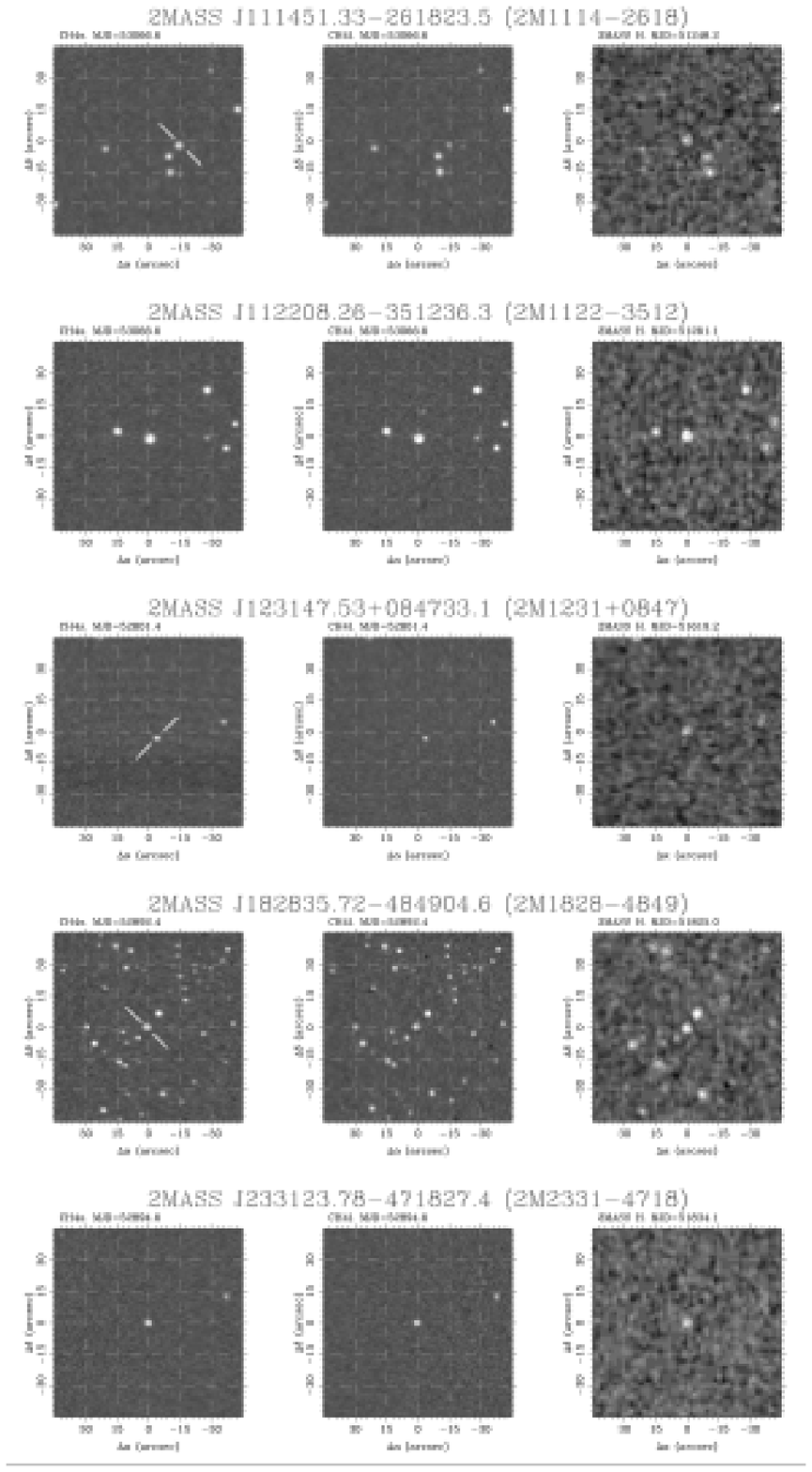}
   \caption{(continued)}
      \label{charts2}
\end{figure*}

\begin{figure*}
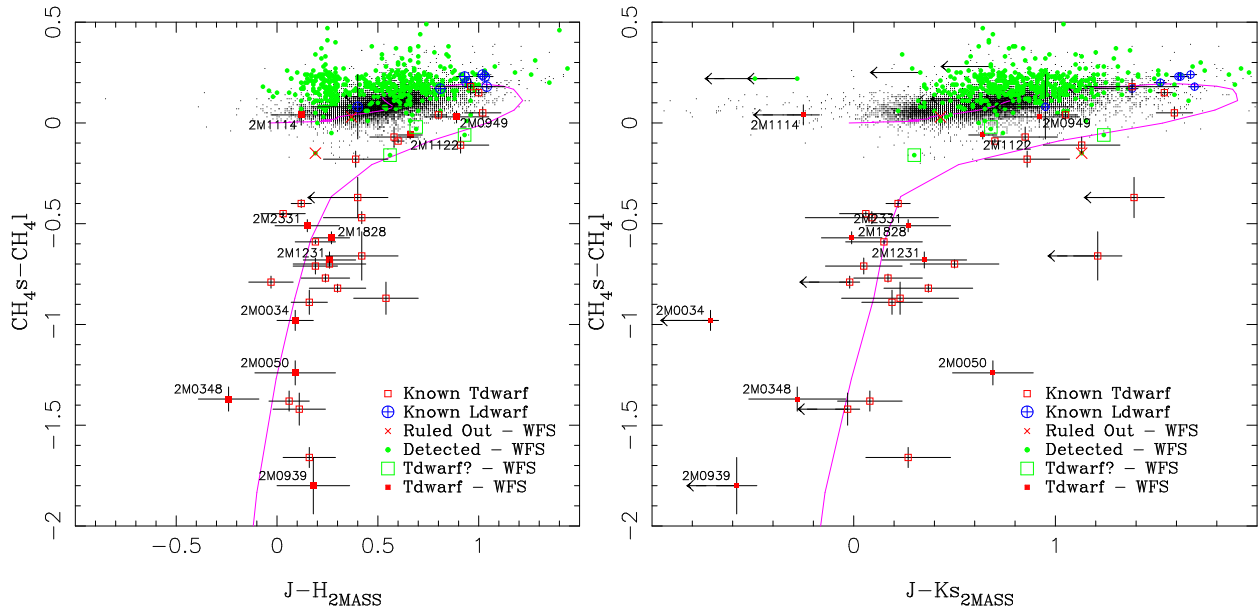

   \centering\includegraphics[width=80mm,angle=270]{f8a.eps}
   \centering\includegraphics[width=80mm,angle=270]{f8b.eps}
   \caption{\footnotesize{(\Msl)$_{Diff}$ plotted against 2MASS J-H and J-Ks for Wide Field T dwarf Search
            (WFTS) candidates observed to date. Objects not detected at Ks or H by 2MASS are
            indicated as colours with upper limits. WFTS T dwarfs
            are plotted as solid squares with labels. WFTS objects which were
            detected, but not indicated as being
            T dwarfs by their \Msl\ photometry, are plotted as solid circles, and
            objects whose \Msl\ photometry suggests a T dwarf status (but not yet
            confirmed by spectra) are highlighted with large open squares, while
            those where spectra have ruled out a T dwarf status are over plotted with
            diagonal crosses. Known
            T dwarfs are shown as open squares (where more than one observation has 
            been taken, they are an average), and known L dwarfs as open circles 
            with crosses.
            The haze of small points underlying the plot shows all background objects
            observed during this program, within 3\arcmin\ of the centre of the 
            IRIS2 field-of-view, and with 2MASS 
            uncertainties of $<$0.05\,mag.}}
      \label{haze}
\end{figure*}

Figure \ref{haze} summarises the results of these differential methane observations.
It plots \Msl$_\mathrm{Diff}$ against 2MASS J-H for (1) a sample of known T and L dwarfs;
(2) the T dwarfs identified from the WFTS; (3) objects detected
by the WFTS methane imaging, but not T dwarfs; (4) a few objects whose methane 
photometry
is not decisive and which need additional spectroscopy to clarify their status; and (5) underlying
these data is the ``haze'' of background objects observed in these 508 IRIS2 observations
with 2MASS photometric uncertainties $<$ 0.05\,mag, and within 3\arcmin\ of
the IRIS2 field centre. The \Msl\ versus (J-H)$_\mathrm{2MASS}$ sequence of 
Table \ref{meth_jhk_table} is also plotted. The T dwarf differential photometry from this figure is also
summarised in Table \ref{diff_table}. 

These results reinforce the power
of methane imaging -- we have reduced the need for spectroscopic confirmation,
from hundreds of objects to  fifteen. It should be noted, however, that
while the methane imaging technique is quite clean for late T dwarfs (of the seven
WFTS candidates imaged in methane and found to have \Msl\ $< -0.5$, all are
spectroscopically found to be T dwarfs), it becomes progressively less so for
early T dwarfs (both the ``ruled out by spectroscopy'' methane candidates have 
\Msl\ $> -0.3$). For early T dwarfs the large number of background objects with
neutral \Msl\ colours, leads to photometric and differential zero-point
uncertainties scattering increasing numbers of background objects into regions where
their colours mimic those of early T dwarfs.  Similarly, the rate at which T dwarfs
could be  ``missed'' by our methane imaging is a strong function of spectral
type. As Table \ref{methdata} indicates, we easily re-detect based on \Msl\ colour alone,
all eleven known T dwarfs later than T4, while for types earlier than T2 position
information in addition to \Msl\ colour is necessary to identify the T dwarf.
A detailed
analysis of all the WFTS candidates observed (required to examine the
efficiency of methane imaging as a function of spectral type, and so
produce selection functions  and luminosity functions from our survey)
is postponed to a subsequent paper when the entire WFTS results will be 
presented and analysed.

A notable feature of Fig. \ref{haze} is that the bluest objects in J-H lie systematically
below the \Msl-(J-H)$_\mathrm{2MASS}$ sequence. This is an artefact of the
initial colour criteria adopted for selection from 2MASS, the 0.1-0.2\,mag uncertainties
on these colours, and the relative numbers of intrinsically red and blue stars
in the 2MASS database. Figure 1 of \citet{bu2003c} shows how the selection used
by the WFTS leads to peaks in the number
of objects selected at (J-H)$_\mathrm{2MASS} \approx 0.7$ (equivalent to mid- to late-M dwarfs) and 
(J-H)$_\mathrm{2MASS} \approx 0.3$ (equivalent to early- to mid-G dwarfs). However, M dwarfs hugely outweigh
G dwarfs in the 2MASS database, meaning that photometric uncertainties scatter
many more actual M dwarfs into the G dwarf ``peak'' than there are real G dwarfs in that peak.
This is reflected in Fig. \ref{haze} where we see that most of the objects in the G dwarf ``peak''
at (J-H)$_\mathrm{2MASS} \approx 0.3$ have \Msl\ colour appropriate for M dwarfs, not G dwarfs.

Proper motions have been estimated for all the T dwarfs listed in Table \ref{diff_table}.
These were calculated by: projecting the 2MASS data in each field into a tangent plane
centred on the 2MASS target catalogue coordinate; using the {\tt match}$^{\ref{match}}$ code to
determine a linear transformation for each astrometrically corrected IRIS2 image to this
tangent plane; and estimating the motion between the 
observation epoch in the 2MASS catalogue and the IRIS2 observation epoch. Typical astrometric uncertainties
due to this process are in the range 0.1-0.2\arcsec, and typical epoch differences are
4-5\,yr. Many of the T dwarfs listed in Table \ref{diff_table} will be at distances of
7-20\,pc, and so will have annual parallax  motions of 0.125-0.05\arcsec, which we have not
attempted to control. These will introduce proper motion uncertainties of a magnitude
similar to those due to the precision of the astrometric transformation between the two
epochs.

\section{Spectroscopy}

\subsection{IRIS2 Spectroscopy}
\label{spectra}

Spectroscopic observations were carried out with IRIS2 on the nights
of 2003 July 12-14, 2004 January 8 and 2004 March 3-9 (UT). 
The targets observed  are summarised in Table \ref{spectra_logs}.
Observations
were obtained with a 1\arcsec\ slit (corresponding to 2.2 pixels on
the detector) in one of three configurations.
\begin{itemize}
\item[Jl] : (Jlong) a 240 line/mm transmission grating bonded to a 
sapphire prism (``Sapphire240'')
used in the collimated beam of IRIS2, together with a spectroscopic blocking
filter denoted ``Jl'' (which transmits $\lambda$=1100-1350\,nm) to deliver 
0.232\,nm per pixel spectra over the wavelength range 1105.2-1343.4\,nm on the detector, with a resolution of 2.3\,pixels, corresponding
to $\lambda/\Delta\lambda=R\approx 2290-2250$.
\item[Hs] : (Hspect) a 316 line/mm transmission grating bonded to a sapphire prism (``Sapphire316'')
used with a spectroscopic blocking
filter denoted ``Hs'' (which transmits $\lambda$=1443-1824\,nm) to deliver 0.346\,nm per pixel spectra over the
wavelength range 1485-1795\,nm (this wavelength range being primarily
determined by terrestrial water vapour transmission), 
with a resolution of 2.3\,pixels, corresponding to $\lambda/\Delta\lambda=R\approx 2050-2150$.
\item[H] : the same configuration as Hs, but with the MKO H filter used as a spectroscopic
blocker. In this mode, the filter bandpass ($\lambda$=1485-1795\,nm) sets the spectral
format recorded (which is nonetheless very similar to that delivered by Hs).
\end{itemize}
Wavelength calibrations were obtained with a Xe lamp, which enables
a third-order polynomial dispersion fit to be obtained with rms scatter
about the fit of better than $1/20^{\mathrm{th}}$ of a pixel.

The spectra were acquired by nodding the telescope between two beams (``A'' and ``B'') in either AB or
ABBA combinations,
with observations at each beam of 300s-600s. For spectroscopic
observing the detector was operated in Multiple Read Mode (MRM), with typically
60 non-destructive reads being acquired over the course of the exposure, and the final
image being obtained from a least-squares fit to each pixel. This mode delivers an effective
read noise in the final image of less than 5 electrons.

Data reduction followed a more-or-less standard procedure for infrared spectrographs.
The first step for all images was to create a Poisson uncertainty image based on
the known gain, read-noise and dark current of IRIS2.
The astrometric distortion present in IRIS2 images (see above), delivers spectra 
with significant curvature (i.e. arc and sky lines are  curved on the
IRIS2 focal plane). For each spectroscopic configuration, arc lines were used
to construct a correction which removes this curvature using the Figaro {\tt sdist} and 
{\tt cdist} codes. These were applied to flat-field images (obtained by subtracting images of
the AAT dome obtained when illuminated with a quartz-halogen lamp, from images obtained with
the lamp turned off) to produce a ``straightened'' flat field image. The spectral response
of this flat-field was then determined by collapsing this image in the spatial direction,
and making a smooth spline fit through it. By then growing this smooth spectral response
into an image, and reversing the distortion correction, we obtained an image which
could normalise the raw flat-field images, and provide a true ``pixel-to-pixel''
spectroscopic flat-field.

All images were flattened with these flat fields, and then
sky subtracted with the relevant exposures in the AB or ABBA observing pattern.
The IRIS2 astrometric distortion was then removed, delivering spectra with constant
wavelength at all columns on the detector, with the majority of the sky lines
subtracted (though changes in the strength of the near-infrared night sky lines
always result in some residual differences being present in these sky-subtracted images).
The tilt of the dispersed spectra relative to the detector rows ($<$0.5\arcdeg)
was then removed.

The observed targets were identified in these processed images, and used to define
windows for a further sky subtraction pass (by fitting along the columns
of the detector), followed by flagging of cosmic rays by hand, and then
optimal extraction. At each step of this process,
the initial photon-counting errors are propagated, so that the final
result is a spectra with meaningful uncertainties. Wavelength calibration was obtained
by passing Xe arc lamp images through the same process, and extracting them from the
same detector regions as each object.

The same procedures were used to reduce observations of A,F and G stars
obtained either immediately before, or immediately after our target
star, in the same  instrument configurations. These were used to correct
the effects of terrestrial atmospheric absorption for each target, assuming
a blackbody flux distribution for each terrestrial correction star
based on their spectral type. The strong hydrogen absorption lines in the H-band spectra
of A type stars were found to make them essentially useless for this purpose,
so these spectra were discarded. These terrestrial correction stars not only enable us
to remove the effects of atmospheric absorption, but also to place our
final spectra on a meaningful F$_\lambda$ flux scale -- though with arbitrary
zero-point calibration uncertainty due to the slit losses when observing
the terrestrial calibration star.

For purposes of spectral typing, the main features of interest are broad,
so we have binned our spectra from their full $R \approx 2200$ resolution to
$R \approx 370$ (i.e. binned by a factor of 6). The binning weighted
each pixel by the inverse square of its uncertainty, resulting in pixels lying
on strong night sky lines receiving low weight in the final binning -- effectively
``software night sky suppression''. Tests with our spectra show that this procedure
can deliver spectra with a given signal-to-noise ratio with an efficiency
about 1.4-1.7 times faster than is possible by simply observing at $R\approx 370$.

\begin{table}
\begin{center}
  \footnotesize
  \caption{IRIS2 Spectroscopic Observing\label{spectra_logs}}
  \begin{tabular}{@{}llccc@{}}
  \colrule\colrule
   Night        & Object & Formats$^a$ & Exposures & Seeing \\
                &        &             & (s)       & (\arcsec) \\
  \colrule
2004 Jan 08     & 2M0050-3322 & Jl,H  & 4$\times$600s,4$\times$600s & 1.5\\
2004 Jan 08     & SD0423-0114 & Jl,H  & 4$\times$300s,4$\times$300s & 1.9\\
2004 Jan 08     & 2M0559-1404 & Jl,H  & 4$\times$300s,4$\times$300s & 2.0\\
2004 Mar 03     & 2M1114-2618 & Jl,Hs & 4$\times$600s,4$\times$600s & 2.4\\
2004 Mar 04     & 2M1122-3512 & Jl,Hs & 4$\times$300s,4$\times$300s & 1.9\\
2004 Mar 04     & 2M1114-2618 & Jl,Hs & 4$\times$300s,4$\times$300s & 1.7\\
2004 Mar 08     & 2M0949-1545 & Jl,Hs & 2$\times$600s,2$\times$600s & 2.4\\
2004 Mar 09     & 2M0939-2448 & Hs    & 2$\times$600s               & 2.2\\
  \colrule
\end{tabular}
\tablenotetext{a}{ See Section \ref{spectra} for wavelength coverage and resolutions of these
formats}
\end{center}
\end{table}

\subsection{Analysis}
\label{spectral_classification}

To place our spectra in a spectral typing sequence (and to derive uniform types
for all the T-dwarfs observed in our methane filters), we 
numerically quantify the large-scale appearance of these using near-infrared
spectroscopic indices. Numerous authors 
(following on from the initial work of \citealt{jones1994,jones1996})
have shown that near-infrared spectra offer multiple features which can be
used to classify ultra-cool dwarfs. The use of  near-infrared spectroscopic indices to 
quantify the strength of broad molecular features in ultra-cool dwarfs 
was pioneered by \citet{delfosse97} (see also \citealt{tinney98,delfosse99}), 
and has since been extended throughout the M and L spectral types
by \citet{reid2001a}, who used the optical M and L dwarf spectral classification
system of \citet{ki2000} as their fundamental standards, and derived
calibrations to transfer infrared spectral indices onto these optical L types.
\citet{testi2001} have continued in this vein for a sample of L dwarfs observed
in the near infrared at very low resolution.
For T dwarfs, near-infrared spectra {\em define} the spectral type. 
Two separate infrared
typing schemes were initially developed by \citet{bu2002a} and 
\citet{ge2002} (the latter including additionally a near-infrared classification of L dwarfs). 
Both are based on spectroscopic indices which are
sensitive to the strength of H$_2$O and CH$_4$ molecular absorption.
A ``hybrid'' scheme was proposed by \citet{bu2003a} to bring these two schemes 
into alignment. Preliminary spectral indices and spectroscopic standards for this 
hybrid scheme were presented by \citet{bu2004a}; the final version of this
scheme is imminent \citep{bu2005}.

To classify the T dwarfs identified by methane-band imaging, we used the H$_2$O$^J$, 
CH$_4^J$, H$_2$O$^H$, and CH$_4^H$ indices of \citet{bu2005}. (These are the same
as those in  \citealt{bu2004a}, but with the numerator of the CH$_4^J$ index 
shifted redward by 0.015$\mu$m). We have used publicly available
L dwarf \citep{kn2004}\footnote{\tt \url{http://www.jach.hawaii.edu/\%7Eskl/LTdata.html}}
and T dwarf\footnote{\tt \url{http://research.amnh.org/\%7Eadam/tdwarf}} 
spectra to calibrate these indices (Table \ref{standard_types}). 
These are monotonic in the H$_2$O indices for
L and T dwarfs, and monotonic in the CH$_4$ indices for objects later than L6.
They therefore enable a spectral type to be estimated
from each index for each T dwarf, with the final spectral
type being adopted as the mean of these values (rounded to the nearest 
half sub-type). 

\begin{figure*}
   \centering\includegraphics[angle=-90, width=160mm]{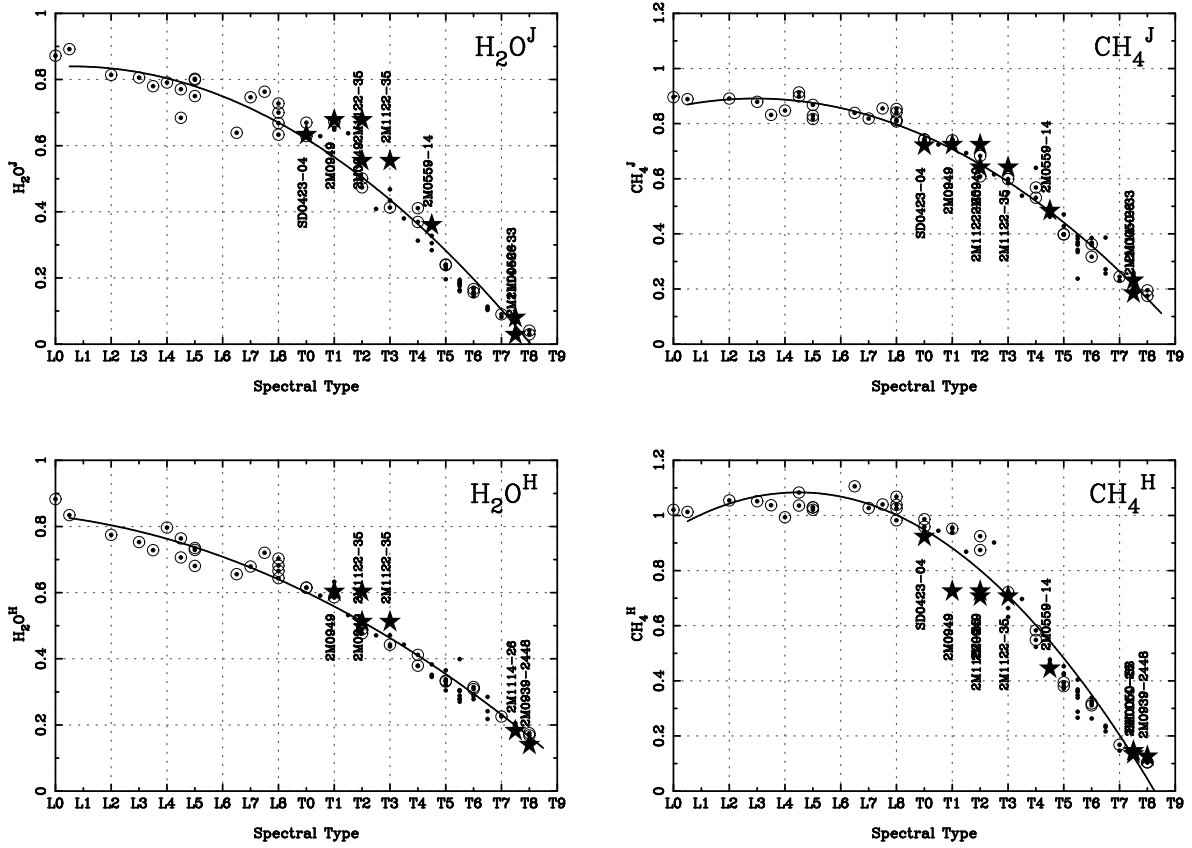}
   \caption{Spectroscopic classification indices  for both our
            fundamental calibration objects (circles with solid dots), 
            other known T dwarfs with ``hybrid'' T types (solid dots), 
            and objects observed with IRIS2 (stars).}
      \label{indices}
\end{figure*}

\begin{figure*}
   \centering\includegraphics[width=140mm]{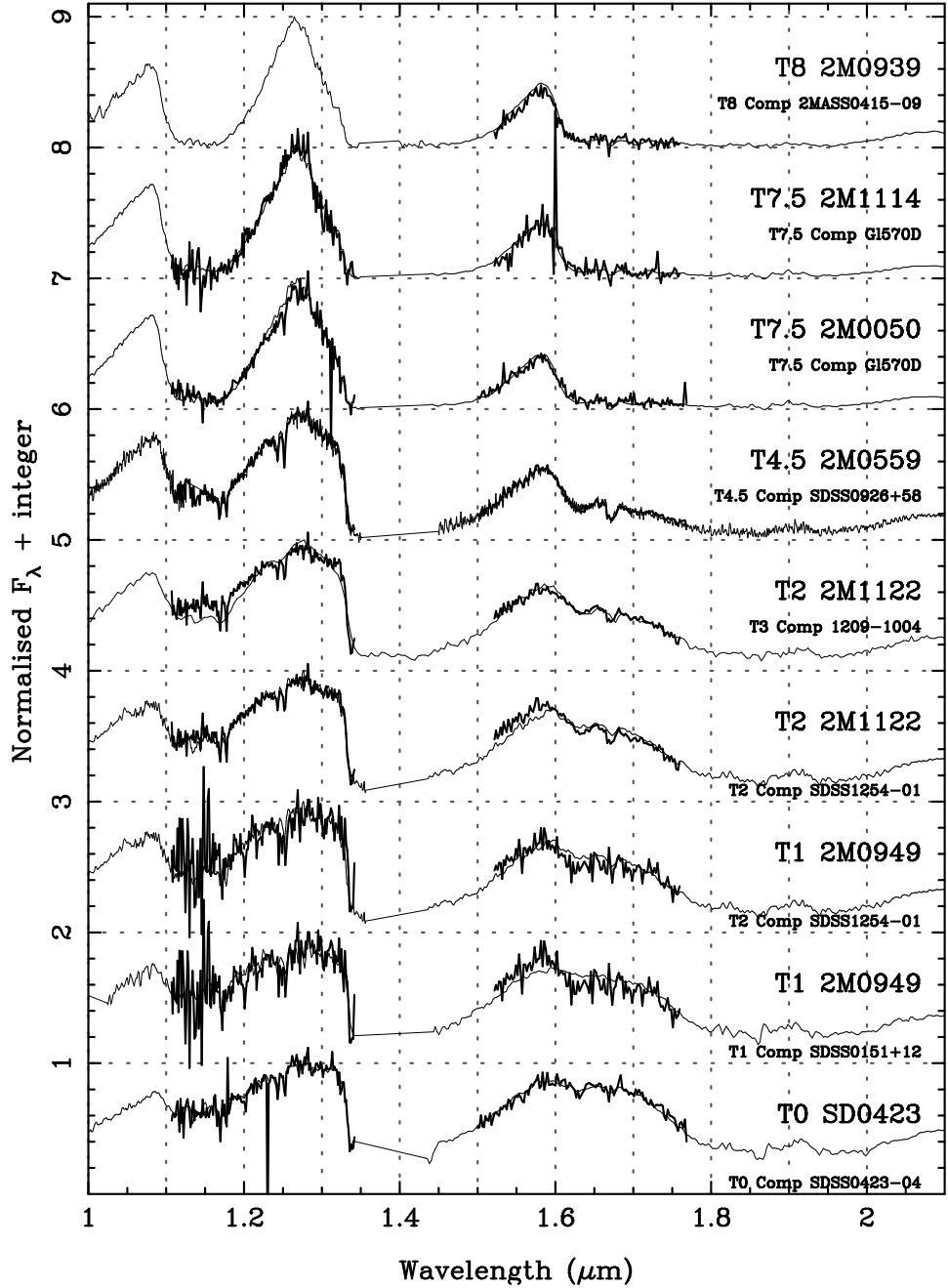}
   \caption{ IRIS2 T dwarf spectra  (thick lines) overplotted on 
             T dwarf comparison spectra (thin lines -- see 
             Section \ref{spectral_classification}). The spectral types
             assigned to the IRIS2 spectra are from Table \ref{specresults}.
             The adopted comparison spectral types are  from the 
             final hybrid classification scheme of \citet{bu2005}. 
           }
      \label{spectrafig}
\end{figure*}

These indices and the derived spectral
types are summarised in Table \ref{specresults} for the T dwarfs observed with IRIS2. 
The uncertainties listed on these
types are the standard error in the mean. These types were then used to ``order'' our 
spectra in a sequence with comparison
spectra -- both our spectra (thick lines) and comparison spectra (underlying thin
lines) are shown in 
Figure \ref{spectrafig} -- each segment of IRIS2 spectrum being normalised to the comparison
spectrum on which it is plotted. The spectra classifications derived are consistent between the
H- and J-band spectra for all the new T dwarfs, except for 2M0949-1545 and 2M1122-3512.
For these objects the 
the spectral type indicated by the H-band spectra are somewhat later than those indicated by the J-band spectra.
This could be due to the effects of unresolved
binarity, or to an underlying structure in the CH$_4^{\mathrm H}$ index calibration, which is not adequately parametrised by the
polynomial fits shown in Fig. \ref{indices}. In either case, the spectral types for these objects should
be regarded as somewhat more uncertain than for the other T dwarfs presented.

\begin{table}
  \begin{center}
  \caption{Spectroscopic Comparison Objects\label{standard_types}}
  \begin{tabular}{@{}lc@{}}
  \colrule\colrule
   SpT        & Object \\
  \colrule
L0   & 2MASS J03454316+2540233                 \\
L0.5 & 2MASS J07464256+2000321                 \\
L3   & DENIS-P J1058.7-1548,    2MASS J15074769-1627386 \\
L3.5 & 2MASS J00361617+1821104                 \\
L4.5 & 2MASS J00283943+1501418, 2MASS J22244381-0158521 \\
L5   & DENIS-P J1228.2-1547AB                  \\
L6.5 & 2MASS J22443167+2043433                 \\
L7   & DENIS-P J0205.4-1159                    \\
L7.5 & 2MASS J08251968+2115521                 \\
L8   & 2MASS J03105986+1648155, 2MASS J03284265+2302051 \\
L8   & 2MASS J15232263+3014562, 2MASS J16322911+1904407 \\
T0   & SDSSp J042348.57-041403.5               \\
T1   & SDSSp J015141.69+124429.6               \\
T2   & SDSSp J125453.90-012247.4               \\
T3   & 2MASS J12095613-1004008                 \\
T4   & 2MASS J22541892+3123498                 \\
T5   & 2MASS J15031961+2525196                 \\
T6   & SDSSp J162414.37+002915.6               \\
T7   & 2MASS J07271824+1710012                 \\
T8   & 2MASS J04151954-0935066                 \\
 \colrule
\end{tabular}
\tablecomments{Adopted types from \citet{ki2000} for L dwarfs 
and \citet{bu2005} for T dwarfs.}
\end{center}
\end{table}

\begin{table*}
\begin{center}
  \caption{Spectroscopic Spectral Typing Results.
           \label{specresults}}
  \begin{tabular}{@{}lccccccc@{}}
  \colrule\colrule
  Object     &  H$_2$O$^\mathrm{J}$ 
                             & CH$_4^\mathrm{J}$ 
                                             & H$_2$O$^\mathrm{H}$ 
                                                             & CH$_4^\mathrm{H}$  
                                                                             &$<$SpT$_\mathrm{spec}>$ 
                                                                                     &SpT$_\mathrm{hyb}$ 
                                                                                          & $<$SpT$_{\mathrm{CH}_4}>$ \\
  \colrule
 SD0423-0114 & 0.63$\pm$0.04 & 0.93$\pm$0.03 &               & 0.92$\pm$0.05 & T0.5$\pm$1.0  & T0 & L3-T0 \\
 2M0949-1545 & 0.68$\pm$0.04 & 0.94$\pm$0.03 & 0.60$\pm$0.03 & 0.72$\pm$0.05 & T1.0$\pm$1.0  &    & T0.3$\pm$1.5\\
 2M1122-3512 & 0.55$\pm$0.04 & 0.89$\pm$0.04 & 0.51$\pm$0.03 & 0.71$\pm$0.05 & T2.0$\pm$0.5  &    & T1.5$\pm$1.1\\
 2M0559-1404 & 0.36$\pm$0.04 & 0.79$\pm$0.04 &               & 0.44$\pm$0.05 & T4.5$\pm$0.5  &T4.5& T4.1$\pm$0.4\\
 2M0050-3322 & 0.08$\pm$0.04 & 0.46$\pm$0.04 &               & 0.14$\pm$0.05 & T7.5$\pm$0.5  &    & T7.1$\pm$0.3\\
 2M1114-2618 & 0.03$\pm$0.04 & 0.38$\pm$0.04 & 0.18$\pm$0.03 & 0.13$\pm$0.05 & T7.5$\pm$0.5  &    & T7.5$\pm$0.3\\
 2M0939-2448 &               &               & 0.14$\pm$0.03 & 0.13$\pm$0.05 & T8.0$\pm$0.5  &    & T8.2$\pm$0.3\\ 
  \colrule
  \end{tabular}
\end{center}
\end{table*}

\begin{table*}
\begin{center}
\caption{Distance and tangential velocity estimates for WFTS T dwarfs\label{Vtan}}
  \begin{tabular}{@{}lcrcc@{}}
  \colrule\colrule
  Object Name 
           &SpT$^a$
                  &2MASS J
                                 &d(pc)$^b$
                                               &V$_{tan}$(km\,s$^{-1}$)$^c$ \\
  \colrule
2M0034+0523 & T6.5  & 15.54$\pm$0.04 &   9 &  37$\pm$19 \\	
2M0050-3322 & T7.5  & 15.93$\pm$0.07 &   8 &  55$\pm$13 \\
2M0348-6022 & T7    & 15.32$\pm$0.05 &   7 &  24$\pm$6  \\
2M0939-2448 & T8    & 15.98$\pm$0.10 &   8 &  46$\pm$8  \\	
2M0949-1545 & T1    & 16.15$\pm$0.12 &  20 &  10$\pm$6  \\
2M1114-2618 & T7.5  & 15.86$\pm$0.08 &   7 & 109$\pm$20 \\	
2M1122-3512 & T2    & 15.02$\pm$0.04 &  14 &  19$\pm$5  \\
2M1231+0847 & T5.5  & 15.57$\pm$0.07 &  15 & 114$\pm$24 \\	
2M1828-4849 & T5.5  & 15.18$\pm$0.05 &  12 &  21$\pm$7  \\
2M2331-4718 & T5    & 15.66$\pm$0.07 &  18 &  18$\pm$9  \\
  \colrule
  \end{tabular}
  \tablecomments{$a$ - Spectroscopic spectral type from Table \ref{specresults} (where IRIS2
                  spectra are available) and Table \ref{diff_table} otherwise. $b$ - 
                  Distance estimate using 2MASS J absolute magnitude versus spectral type
                  relation of \citet{tparallax} resulting in typical distances uncertainties 
                  of $\pm$20\%. $c$ - Uncertainty includes 0.36\,mag scatter about absolute magnitude
                  spectral type relation and photometric uncertainties for each T dwarf.}
\end{center}
\end{table*}

\section{Discussion}
\label{discussion}

\subsection{The New T dwarfs}

Five of the WFTS T dwarfs reported in Table \ref{diff_table} (2M0050-3322, 
2M0939-2448, 2M0949-1545, 2M1114-2618 \& 2M1122-3512) are reported for the 
first time by this paper. A further four objects  (2M0034+0523, 2M1231+0847, 
2M1828-4849 \& 2M2331-4718) were confirmed as T dwarfs by our methane imaging
in parallel with the traditional spectroscopic approach, and have already been 
published by \citet{bu2004a,bu2004b}. Three of the five new T dwarfs 
(2M0050-3322, 2M1114-2618 \& 2M0939-2448) are very 
cool objects with spectral types of T7.5 or later. In comparison, the other 
two objects are very early -- at T1 (2M0949-1545) and T2 (2M1122-3512) they 
are the earliest T dwarfs yet found by selection for T dwarfs
from the 2MASS database.

Of the WFTS T dwarfs reported here, 2M0939-2448 shows pronounced evidence in 
Fig. \ref{haze} for the same K-suppressed spectrum previously seen 
in 2M0034+0523 \citep{bu2004a} and 2M0937+2931 \citep{bu2002a} --
both 2M0939-2448 and 2M0034+0523 stand out as having the bluest J-Ks colours 
of any of the T dwarfs plotted (and indeed being only upper limits at Ks in 2MASS). 
Their J-H colours, by contrast are not anomalous 
in any way whatsoever. Note that the anomalously red J-Ks colours of SD0207+0000 
and SD1110+0116 are not physical, but reflect the upper limits for K-band detection 
from 2MASS of these faint SDSS-discovered objects. On the other hand 2M0050-3322 does
seem to be anomalously red for its methane index. 
A detailed discussion of the
possible causes of K-band flux suppression in T dwarfs is postponed to a
forthcoming paper.

Spectro-photometric distance estimates have been derived using the
2MASS J absolute magnitude versus spectral type relation of 
\citet{tparallax}. These result in typical distances uncertainties of $\pm$20\% and
 are presented in Table \ref{Vtan}. They place  all
of these WFTS T dwarfs at distances of between 5 and 20\,pc, making them all
likely
members of the important and well studied ``Nearby Star'' volume-limited sample of objects within
25\,pc of the Sun. They are all ideal
targets for a 4m-class telescope near-infrared parallax program, and we
initiated such a program with IRIS2 on the AAT in 2004. Of these T dwarfs, two (2M1114-2618
and 2M1231+0847) show appreciable proper motions, indicating tangential velocities
$\simgt$100\,\kms.

\subsection{Applications of Methane Imaging}

As outlined in section \ref{background}, methane imaging offers significant
advantages over current techniques for the identification and classification
of T dwarfs. We have shown this is particularly true for   ``winnowing''
these rare objects from a large database like 2MASS, where we could  use
methane imaging in place of the combination of infrared imaging plus optical
imaging plus infrared spectroscopy previously employed. Moreover, the
greatly simplified data processing allows almost instantaneous feed back
and spectroscopic confirmation, if necessary.

\subsubsection{Large Survey Winnowing}

Perhaps the most obvious application for  methane imaging is the winnowing
of forthcoming large surveys for cool brown dwarfs. The UKIRT Infrared Deep 
Sky Survey (UKIDSS;  \citealt{ham03}), for example, will carry out surveys
in the Y, J, H and K passbands totalling a volume sensitivity for cool brown dwarfs
almost 250 times larger than that obtained by 2MASS. While the filters used
in these surveys will be far more suitable for the selection of cool brown
dwarfs than those used by 2MASS \citep{le2005}, it is nonetheless true that
T dwarfs in UKIDSS will represent a similar, tiny fraction 
of the total number of sources ($\simlt 3\times10^{-5}$) as was seen in 2MASS.
Hard won experience shows that when selecting objects this rare from a large
database, the resulting sample is {\em always} dominated by contaminants -- detector
artefacts, ghosts, asteroids, cosmic rays, partially resolved binaries, confused
sources near the survey limits, etc. Such contaminants produce catalogue
sources with arbitrarily odd colours, some of which will mimic T dwarfs.  The class 
of brown dwarfs cooler than T dwarfs (tentatively
known as the  ``Y'' dwarfs, and one of the key science drivers for the UKIDSS
survey) will be even rarer, and candidate samples will be subject to even more 
contamination. Models
for these objects suggest that they will have extremely unusual colours, and will
lie some way from the sequence of foreground and background G-M dwarfs in a J-H versus
Y-J colour-colour diagram -- as much as a magnitude redder in Y-J, and a magnitude bluer in J-H.
The $\pm$0.1-0.2\,mag photometric uncertainties of objects near the survey
magnitude limits will only rarely scatter G-M dwarfs by the 5-10$\sigma$ needed to
move such objects into the Y dwarf selection region. But this will only need
to happen at a rate of once in a billion to still massively outnumber the likely
number of true Y dwarfs in the survey.

Some form of winnowing of candidates is almost inevitably going to be required
before taking H$\approx$18 Y dwarf candidates to an 8m telescope for infrared
spectroscopy. Methane imaging  offers an ideal mechanism for doing this,
since H$\approx$18 can be imaged on a 4m telescope in under an hour, compared to the
several hours required to acquire spectra on an 8m telescope.

\subsubsection{Cluster T dwarfs}

It has been a common-place amongst astronomers for many years that
there must exist a minimum mass for star formation, below which
the mass function ``turns over''. Based initially on
Jeans mass arguments \citep{ll1976}, and later on 
hydrodynamic simulations \citep{boss1983,boss1993} this minimum mass was
placed at $\sim$0.01\,\Msol\ (or $\sim$10\,\Mjup). The similarity of
this value with the minimum mass for deuterium burning at $\sim$\,13\,\Mjup, has 
lead to a 10-13\,\Mjup\ value being widely assumed as the minimum mass 
for star formation (or the maximum mass for planet 
formation\footnote{\tt \url{http://www.ciw.edu/boss/IAU/div3/wgesp/definition.html}}) and so
as a dividing line between stars and planets. Below this mass, compact
astrophysical objects are assumed to form {\em within} planetary disks
via the ``bottom up'' processes of planetary accretion.
Unfortunately, this comforting paradigm has become somewhat more confused of late. 
\begin{enumerate}
\item  \citet{boss2001} has included some magnetic field effects in  simulations
   which show that star formation processes can make compact objects as small as
  1\,\Mjup.
\item Candidate objects with masses which theoretical models suggest have
  masses similar to and smaller than 10\,\Mjup, are being turned up
  by surveys in star forming regions, in an apparent extension of
  ``star'' formation to ``planetary'' masses.
  These potentially ``unbound'' planets represent 
  a challenge and a conundrum
  for the simple paradigm above -- are they brown dwarfs formed via
  star formation processes, stellar embryos ejected from multiple systems before
  they can accrete sufficient matter to become stars \citep{rc2001}, or 
  giant planets ejected from their host stars?
\item Measurements of the brown dwarf mass function in 
  a variety of moderate age clusters ($>$100\,Myr) are finding results in the range
  $\propto$\,m$^{-0.5}$ to $\propto$\,m$^{-0.6}$ \citep{bo2003} suggesting
  a mass function which is continuous with the stellar function,
  but which continues to turn over at lower and lower masses, extending into
  the T dwarf range in some clusters.
\item Radial velocity programs are now detecting more
   massive analogues (extending up to 15\,\Mjup) of the 
   m$\sin$\,$i$=0.1-5\,\Mjup\ exoplanets. The  mass function 
   of these exoplanets is consistent with a single power law
  $\propto$\, [m\,$\sin$\,$i$]$^{-0.8}$ or steeper \citep{mar2004}. This is 
   {\em steeper} again than the cluster brown dwarf mass function.
\end{enumerate}

There is a need to clarify the status of the mass function
in the range where the lines between
planets and brown dwarfs are so blurred. Determining the relative normalisations
between the two mass functions is essentially impossible -- however, we can
attempt to address the {\em shape} of the mass function in this
boundary region. That is try to answer the question
{\em ``Does the star formation mass function turn over between
40 and 5\Mjup?''}. If it does, then we must determine where it does, as this
is critical information needed to refine models of the star formation
process.

Unfortunately, while sky surveys have
been extremely successful at detecting field brown dwarfs,  their
unknown ages make the measurement of masses problematic. Such studies need to target stellar clusters 
of known age in order to estimate  masses.
Unfortunately, clusters close enough to the Sun for brown dwarfs 
to be detectable are
 large on the sky (typically 0.2-10\,sq.deg.), and typically show  low 
contrast against the low latitude background of field stars. 
Moreover, the photospheres of cool brown dwarfs are
phenomenally complex, leading to real difficulties for theoretical
models. This makes traditional colour-magnitude diagrams a  poor
means of selecting cluster members, because uncertainties force
the selection of very large samples in colour-magnitude space, with
poor hit-rates for {\em actual} cluster members. The resultant
imaging samples  then  require substantial spectroscopic winnowing  
on 8-10m telescopes.

T dwarf methane absorption, however,  is so spectrally unique, that
 it provides a potentially powerful tool 
for selecting candidate T dwarfs in clusters. 
\citet{tparallax} have looked at the absolute magnitudes at which
methane may begin to appear in some representative clusters, and 
concluded that in nearby (d$\approx$150\,pc), 10-20\,Myr old 
clusters like IC\,2391 or IC\,2602,
the absolute magnitude for CH$_4$ to become detectable by methane imaging (equivalent to
spectral types T2 and later)  will be M$_H$$\sim$12.5,  
or equivalently H$\sim$18.2. Later T types will appear at
at fainter magnitudes, with the more easily detectable levels of
methane absorption seen in T5 dwarfs appearing around 1.5\,mag fainter.

Using the available T dwarf models 
(see \citealt{tparallax} and references therein), we can
estimate the second-order {\em differential} 
corrections to this first-order methane onset estimate.
If we adopt the absolute magnitude corresponding to an easily
detectable spectral type of T5 as our measure
(hereafter M$_{\mathrm CH4}^{T5}$) , we can estimate 
offsets from the M$_{\mathrm CH4}^{T5} \approx 14$ of field
dwarfs to the M$_{\mathrm CH4}^{T5}$ for  younger clusters,
 as well as the masses to which these M$_{\mathrm CH4}^{T5}$ values correspond. 
For clusters of
age 100, 50, 10 \& 5\,Myr the relevant offsets are found to
be (respectively) $\Delta$M$_{\mathrm CH4}^{T5}$ = -0.7, -1.0, -1.3 \& -1.5 and
the masses to which these M$_{\mathrm CH4}^{T5}$ correspond 
are 0.02, 0.012, 0.008 \& 0.006\,\Msol. 
For any given cluster, therefore,
a survey can be designed using 
CH$_4$ selection to probe mass functions at unprecedentedly low
masses. As an example, the IRIS2 imager is able to detect T5 dwarfs in
the IC\,2391 cluster (d=140\,pc, age=20$\pm$10\,Myr) after just 
8 hours of integration (i.e.. 4 hours in each bandpass),
reaching to a depth of H=21 over an area of 64\,sq.arcmin., 
corresponding to a detection threshold mass of $\approx$10\,\Mjup.

The resulting T dwarf imaging candidates will still need to be followed
up spectroscopically, in order to distinguish young cluster members,
from old field T dwarf contaminants. But it is nonetheless true that
contamination by {\em non-T dwarfs} can be arranged to be small, or
negligible, vastly improving the speed of confirmation, or rejection,
of cluster membership for T dwarfs.

\subsubsection{Cool companions to Known Brown Dwarfs}

Very rare astronomical objects (like T and Y dwarfs) are typically
searched for by surveying very large chunks of the sky.
An alternate approach  is to
rely on the fact that most astronomical objects are clustered, and so to search
for rare objects in the vicinity of a objects of a similar nature. It was exactly such a search
technique that revealed the first T dwarf Gl\,229B \citep{nak1995}. The invariable
difficulty
in such searches is telling the difference between a potentially
interesting companion to the target in question, and the vast majority of
background objects. In the case of Gl\,229B, the companion was revealed by its
common proper-motion. Methane imaging offers a potentially powerful way to identify
cool companions to nearby stars and brown dwarfs, without the need to wait for
the year or more needed to confirm or deny common proper motion. For example,
a few hours integration with the methane filters in IRIS2 can easily reach to
sensitivities adequate to detect methane absorption in objects 3-4 magnitudes
fainter than all the T dwarfs so far identified from 2MASS, making this a potentially
easier way in which to find Y dwarfs than a large blind survey.

\subsubsection{Methane Variability and ``Weather''}

There is now considerable evidence to indicate that many, if not most, brown dwarfs
rotate quite rapidly, with either measured 
\citep{bjl2003,clarke2002a,clarke2002b,enoch2003,gelino2002,koen2003,bjm2001,tt1999}
 or inferred \citep{bj2004,sm2003,moh2003,reid2002,tr1998}
rotation timescales of 1-10\,h. L and T type brown dwarfs are also well known to suffer
greater or lesser degrees of condensate and/or cloud layer formation (see e.g.
\citealt{ack2001,allard2001,bs1999}).
So brown dwarfs have the two ingredients required for the existence of complex
rotationally driven weather patterns on their surface. Unfortunately, despite 
much intensive searching (see references above) detailed evidence for
weather variations in brown dwarfs remains elusive.

However, as the source of the dominant absorption feature in the spectra
of T dwarfs, searches for variability in CH$_4$ absorption (reflecting
an uneven surface coverage at the optical depths where these absorptions
are formed) could be extremely powerful \citep{ti1999}. Jupiter, for example,
 is known to show strong
surface features in the infrared where gaps in its 
methane layers allow the
hotter, lower regions of the photosphere to shine through (see for 
example the images of Jupiter in \citet{gm2000}).
The nearly differential nature of 
methane imaging should make possible significantly higher precisions (i.e. $<$0.01\,mag)
in programs targeting single objects in extended campaigns, than have been achieved in this program. 
In particular, bright T dwarfs can be targeted (or larger telescopes
used) to improve photon-counting uncertainties, observation of a single object for variability
makes the zero-point uncertainty associated with the selection of an ensemble of
background objects moot, and  use of the \Msl\ colour should obviate many of the
second-order differential extinction effects discussed by \citet{bjl2003}.

\section{Conclusion}

We have shown that methane filters can be used to efficiently detect and characterise
T dwarfs. We have provided a procedure for calibrating methane
filter observations onto a \Ms,\Ml\ photometric system using 
2MASS photometry as a calibration system.
Five new T dwarfs discovered as part of the 2MASS Wide Field T dwarf Search
using this technique have been presented. Finally we have discussed potential
further uses for these methane filters including: the winnowing of large survey datasets
for T and Y dwarfs; the detection of very low-mass brown dwarfs (as low as 10\,\Mjup)
in nearby star clusters; the detection of cooler companions to already known brown dwarfs; and,
the detection of variability signatures due to rotating cloud structures in T dwarfs.

\acknowledgements

The authors wish to gratefully acknowledge the Joint Astronomy
Centre, Hawaii, for making their ORAC-DR code available, and
for assisting the AAO in implementing it for IRIS2. This wonderful
data reduction package has {\em significantly} improved the
efficiency with which this observing program was carried out.
We would also like to acknowledge Michael Richmond,
Emmanuel Bertin and Tim Pearson -- the authors of
the {\tt match}, {\tt sextractor} and {\tt PGPLOT} codes which contributed
significantly to this work.
AJB acknowledges  support by the National Aeronautics and Space
Administration (NASA) through the SIRTF Fellowship program. JDK acknowledges the 
support of the Jet Propulsion Laboratory, California Institute of Technology,
which is operated under contract with NASA. CGT joins his US colleagues in
thanking his employer for paying him. 
This publication makes use of data from the Two Micron All Sky
Survey, which is a joint project of the University of Massachusetts and the
Infrared Processing and Analysis Center, funded by NASA and the NSF.
It also makes use of the NASA/IPAC Infrared Science Archive, 
which is operated by the Jet Propulsion Laboratory, California Institute of 
Technology, under contract with the National Aeronautics and Space Administration.




\clearpage
\begin{landscape}
\tabletypesize{\footnotesize}
\begin{deluxetable}{lllcrrrcccc}
\tablecaption{\Msl\ and JHK photometry for A to T dwarfs\label{methdata}}
\tablewidth{0pt}
\tablehead{
\colhead{Object Name} 
            &\colhead{} 
                      &\colhead{SpT}
                            &\colhead{$n$\tablenotemark{a}}
                                 &\colhead{J-H$_\mathrm{MKO}$\tablenotemark{b}} 
                                         &\colhead{H-K$_\mathrm{MKO}$\tablenotemark{b}}
                                                  &\colhead{J-K$_\mathrm{MKO}$\tablenotemark{b}}
                                                         &\colhead{\Msl\tablenotemark{c}}
                                                                        &\colhead{\Msl$_\mathrm{Diff}$\tablenotemark{d}}
                                                                                                             &\colhead{Src SpT\tablenotemark{e}}
                                                                                                                   &\colhead{Src JHK\tablenotemark{f}}
          }
\startdata
{\bf UKIRT FS}\\
FS10        &GD50                      &DA   &-1  &-0.05 &-0.14 &-0.19 &-0.09$\pm$0.06 (1)&-0.11$\pm$0.04 (1)&  0  &  1     \\     
FS20			&EG76                      &DA   &-1  &-0.01 &-0.05 &-0.07 &-0.04$\pm$0.05 (1)&-0.06$\pm$0.02 (1)&  0  &  1     \\     
FS34        &EG141                     &DA   &-1  &-0.04 &-0.09 &-0.12 & 0.00$\pm$0.03 (2)&-0.02$\pm$0.01 (2)&  0  &  1     \\     
FS7         &SA94-242                  &A2   &2   & 0.12 & 0.01 & 0.13 & 0.01$\pm$0.02 (3)&+0.06$\pm$0.02 (3)&  1  &  1     \\     
FS11        &SA96-83                   &A3   &3   & 0.07 & 0.03 & 0.09 &-0.01$\pm$0.02 (3)&+0.04$\pm$0.01 (3)&  1  &  1     \\     
FS2         &SA92-342                  &F5   &15  & 0.19 & 0.03 & 0.22 &-0.01$\pm$0.03 (1)&-0.03$\pm$0.01 (1)&  1  &  1     \\     
FS4         &SA93-317                  &F5   &15  & 0.23 & 0.03 & 0.27 &-0.01$\pm$0.03 (1)&+0.01$\pm$0.01 (1)&  1  &  1     \\     
FS18        &SA100-280                 &F8   &18  & 0.24 & 0.05 & 0.29 & 0.02$\pm$0.02 (2)&+0.07$\pm$0.01 (2)&  1  &  1     \\     
FS137			&                          &G1   &21  & 0.25 & 0.05 & 0.30 & 0.03$\pm$0.02 (1)&+0.02$\pm$0.01 (2)&  1  &  1     \\     
FS16        &                          &G1   &21  & 0.29 & 0.06 & 0.34 & 0.06$\pm$0.01 (1)&+0.03$\pm$0.01 (1)&  1  &  1     \\     
FS13        &SA97-249                  &G4   &24  & 0.31 & 0.05 & 0.36 & 0.00$\pm$0.01 (3)&+0.04$\pm$0.01 (3)&  1  &  1     \\     
FS135			&                          &G5   &25  & 0.30 & 0.06 & 0.35 & 0.03$\pm$0.01 (1)&+0.06$\pm$0.01 (1)&  1  &  1     \\     
FS136			&                          &K2   &31  & 0.52 & 0.12 & 0.64 & 0.11$\pm$0.01 (1)&+0.09$\pm$0.02 (1)&  1  &  1     \\     
FS124       &LHS254                    &M5   &40  & 0.40 & 0.34 & 0.74 & 0.13$\pm$0.03 (1)&+0.18$\pm$0.01 (1)&  2  &  1     \\     
FS128			&                          &M5   &40  & 0.55 & 0.38 & 0.93 & 0.14$\pm$0.02 (1)&+0.16$\pm$0.01 (1)&  2  &  1     \\     
FS129			&LHS2397a                  &M8(+L7)&43  & 0.63 & 0.54 & 1.17 & 0.16$\pm$0.01 (4)&+0.21$\pm$0.01 (4)&  2  &  1     \\[2pt] 
{\bf L dwarfs}\\
2M1045-0149	&2MASS J10452400-0149576   &L1
                                             &46  & 0.65 & 0.63 & 1.27 & 0.18$\pm$0.03 (1)&+0.17$\pm$0.01 (1)&  7  &  3    \\     
Kelu-1		&2MASS J13054019-2541059   &L2   &47  & 0.78 & 0.67 & 1.45 & 0.20$\pm$0.02 (1)&+0.23$\pm$0.02 (1)&  3  &  2    \\     
DEN1058-1548&DENIS-P J1058.7-1548      &L3   &48  & 0.84 & 0.74 & 1.57 & 0.22$\pm$0.02 (1)&+0.23$\pm$0.02 (1)&  3  &  2    \\     
DEN1539-0520&DENIS-P J153941.96-052042.4&L4
                                             &49  & 0.69 & 0.57 & 1.26 & 0.14$\pm$0.02 (1)&+0.08$\pm$0.16 (1)&  4  &  3    \\     
DEN1228-1547&DENIS-P J1228.2-1547AB    &L5   &50  & 0.88 & 0.69 & 1.57 & 0.20$\pm$0.03 (1)&+0.23$\pm$0.02 (1)&  3  &  2    \\     
2M1507-1627	&2MASS J15074769-1627386   &L5
                                             &50  & 0.75 & 0.67 & 1.42 & 0.16$\pm$0.02 (1)&+0.20$\pm$0.01 (1)&  3  &  3    \\     
DEN0255-4700\tablenotemark{g}
            &DENIS-P J0255.0-4700      &L8   &53  & 0.83 & 0.74 & 1.57 & 0.12$\pm$0.03 (1)&+0.18$\pm$0.01 (1)&  3  &  4    \\     
{\bf T dwarfs}\\
 SD0423-0414 &SDSSp J042348.57-041403.5 &T0   &54  & 0.79 & 0.55 & 1.34 & 0.08$\pm$0.03 (2)&+0.15$\pm$0.02 (4)&  5  &  3     \\     
 SD1207+0244 &SDSSp J120747.17+024424.8 &T0.5 &54.5& 0.75 & 0.47 & 1.22 &-0.05$\pm$0.05 (1)&+0.05$\pm$0.03 (1)&  5  &  3    \\[2pt]    
 SD1254-0122 &SDSSp J125453.90-012247.4 &T2   &56  & 0.53 & 0.29 & 0.82 &-0.03$\pm$0.06 (1)&+0.04$\pm$0.02 (1)&  5  &  3     \\     
 SD1021-0304 &SDSSp J102109.69-030420.1 &T3.5 &57.5& 0.47 & 0.15 & 0.62 &-0.30$\pm$0.07 (1)&-0.11$\pm$0.04 (4)&  5  &  3     \\     
 SD0207+0000 &SDSSp J020742.83+000056.2 &T4   &58  & 0.01 &-0.03 & 0.04 &-0.29$\pm$0.04 (1)&-0.37$\pm$0.10 (1)&  5  &  3     \\     
 2M0559-1404 &2MASS J05591914-1404488   &T4.5 &58.5&-0.16 &-0.07 &-0.09 &-0.43$\pm$0.03 (1)&-0.40$\pm$0.02 (3)&  5  &  3     \\     
 2M0516-0445 &2MASS J05160945-0445499   &T5.5 &59.5&-0.06 & 0.16 & 0.10 &-0.79$\pm$0.03 (1)&-0.70$\pm$0.02 (3)&  6  &  4    \\     
 SD1110+011  &SDSSp J111010.01+011613.1 &T5.5 &59.5&-0.10 & 0.17 & 0.07 &-0.69$\pm$0.20 (2)&-0.66$\pm$0.12 (2)&  5  &  3     \\     
 2M2356-1553 &2MASS J23565477-1553111   &T5.5 &59.5&-0.22 &-0.03 &-0.25 &-0.93$\pm$0.04 (1)&-0.71$\pm$0.05 (4)&  5  &  3     \\     
 2M0243-2453 &2MASS J02431371-2453298   &T6   &60  &-0.26 & 0.05 &-0.21 &-0.79$\pm$0.05 (1)&-0.77$\pm$0.02 (3)&  5  &  3     \\     
 2M2228-4310 &2MASS J22282889-4310262   &T6   &60  &-0.05 &-0.03 &-0.08 &-0.89$\pm$0.03 (1)&-0.82$\pm$0.02 (1)&  5  &  4    \\     
 SD1346-0031 &SDSSp J134646.45-003150.4 &T6.5 &60.5&-0.35 & 0.11 &-0.24 &-1.00$\pm$0.05 (1)&-0.87$\pm$0.08 (1)&  5  &  3     \\     
 2M1217-0311 &2MASS J12171110-0311131   &T7   &61  &-0.42 & 0.06 &-0.36 &-1.44$\pm$0.08 (1)&-1.42$\pm$0.08 (1)&  5  &  3     \\     
 Gl570d		 &2MASS J14571496-2121477   &T7.5 &61.5&-0.46 &-0.24 &-0.70 &-1.54$\pm$0.04 (1)&-1.38$\pm$0.05 (4)&  5  &  3     \\   
 2M0415-0935 &2MASS J04151954-0935066   &T8   &62  &-0.51 &-0.38 &-0.13 &-1.69$\pm$0.34 (1)&-1.66$\pm$0.05 (3)&  5  &  3     \\     
 \enddata
\tablenotetext{a}{ $n$ is the spectral subtype plus a constant for A-T dwarfs: 0 for an A dwarf; 10 for F; 20 for G;
29 for K; 35 for M; 45 for L and 54 for T. $n$ is set to the placeholder value of -1 for white dwarfs.}
\tablenotetext{b}{ J-H,H-K and J-K on the MKO photometric system. 
Typical uncertainties are $<$0.01\,mag for UKIRT FS standards and $<$0.05\,mag for the remaining
objects.}
\tablenotetext{c}{ The number of observations averaged to provide the quoted result are indicated in parentheses.}
\tablenotetext{d}{ See text Section \ref{differential} for definition of \Msl$_\mathrm{Diff}$. The number of observations averaged to provide the quoted result are indicated in parentheses.}
\tablenotetext{e}{ 0 - \citet{ms1999}; 1 - \citet{ha2001}; 2 - \citet{le2002}; 3 - \citet{ki2000} ; 4 - Spectral 
                   type provided by D.Kirkpatrick (private communication) and determined using
                   optical spectroscopy to estimate spectral types on the \citet{ki2000} system; 
                   5 - \citet{bu2005}; 
                   6 - \citet{bu2003c}; 7 - \citet{gizis2002}}
\tablenotetext{f}{ 1 -  UKIRT MKO photometric standards (see Section \ref{absolute}); 2 - \citet{le2002} ;
3 - \citet{kn2004}; 4 - converted 2MASS All Sky data into the MKO system using the conversion relations of \citet{st2004}}
\tablenotetext{g}{The conversions from 2MASS to MKO photometry of \citet{st2004} are essentially independent of
spectral typing system for early-to-mid L dwarfs and T dwarfs, but are less so for late-type L dwarfs.
In the case of DENIS-P J0255.0-4700 this could lead to an additional uncertainty in its MKO
colours of $<$0.03\,mag over and above the 0.05\,mag relevant for the other L and T dwarfs so
converted.}
\end{deluxetable}
\clearpage
\end{landscape}

%
%
%

\clearpage
\begin{landscape}
\tabletypesize{\footnotesize}
\begin{deluxetable}{lrcrrrcccccc}
\tablecaption{\Msl\ photometry, proper motions and spectral types for T dwarfs\label{diff_table}}
\tablewidth{0pt}
\tablehead{
\colhead{Short Name} 
            &\colhead{Full Name} 
                                      &\colhead{SpT\tablenotemark{a}}
                                                   &\colhead{CH$_4$ SpT\tablenotemark{b}}
                                                                   &\colhead{\Msl$_\mathrm{Diff}$\tablenotemark{c}}
                                                                                        &\colhead{J-H$_\mathrm{2M}$\tablenotemark{d}} 
                                                                                                         &\colhead{H-Ks$_\mathrm{2M}$\tablenotemark{d}}
                                                                                                                          &\colhead{J-Ks$_\mathrm{2M}$\tablenotemark{d}}
                                                                                                                                                 &\colhead{$\mu$\tablenotemark{e}} 
                                                                                                                                                            &\colhead{$\theta$\tablenotemark{e}}
                                                                                                                                                                           &\colhead{Src\tablenotemark{f}}
          }
\startdata
{\bf WFTS}  & \\
2M0034+0523 &2MASS J00345157+0523050   & T6.5      & T6.5$\pm$0.3  &$-$0.98$\pm$0.05 (1)&$+$0.09$\pm$0.09&$-$0.80$\pm$0.08&$-$0.71$\pm$0.04:&  0.8$\pm$0.3   &   85$\pm$15  & 1 \\
2M0050-3322 &2MASS J00501994-3322402   & \nodata   & T7.1$\pm$0.3  &$-$1.24$\pm$0.06 (1)&$+$0.09$\pm$0.20&$+$0.60$\pm$0.27&$+$0.69$\pm$0.20&  1.5$\pm$0.1   &   53$\pm$5   & 3 \\
2M0348-6022 &2MASS J03480772-6022270   & T7        & T7.4$\pm$0.3  &$-$1.37$\pm$0.06 (1)&$-$0.24$\pm$0.15&$-$0.04$\pm$0.27&$-$0.28$\pm$0.24& 0.76$\pm$0.06  &  196$\pm$6   & 1 \\
2M0939-2448 &2MASS J09393548-2448279   & \nodata   & T8.2$\pm$0.3  &$-$1.80$\pm$0.14 (1)&$+$0.18$\pm$0.18&$-$0.76$\pm$0.15&$-$0.58$\pm$0.10:& 1.15$\pm$0.06  &  155$\pm$1   & 3 \\
2M0949-1545 &2MASS J09490860-1545485   & \nodata   & T0.3$\pm$1.5  &$+$0.03$\pm$0.03 (2)&$+$0.89$\pm$0.16&$+$0.04$\pm$0.20&$+$0.92$\pm$0.20& 0.10$\pm$0.04  &  271$\pm$34  & 3 \\
2M1114-2618 &2MASS J11145133-2618235   & \nodata   & T7.5$\pm$0.3  &$-$1.40$\pm$0.07 (1)&$+$0.12$\pm$0.15&$-$0.38$\pm$0.12&$-$0.25$\pm$0.08:& 3.05$\pm$0.04  &  263.2$\pm$0.8 & 3 \\
2M1122-3512 &2MASS J11220826-3512363   & \nodata   & T1.5$\pm$1.1  &$-$0.06$\pm$0.02 (1)&$+$0.66$\pm$0.06&$-$0.02$\pm$0.08&$+$0.64$\pm$0.07& 0.29$\pm$0.03  &  211$\pm$8   & 3 \\
2M1231+0847 &2MASS J12314753+0847331   & T5.5      & T5.5$\pm$0.4  &$-$0.68$\pm$0.04 (1)&$+$0.26$\pm$0.13&$+$0.09$\pm$0.22&$+$0.35$\pm$0.21& 1.61$\pm$0.07  &  227$\pm$4   & 1 \\
2M1828-4849 &2MASS J18283572-4849046   & T5.5      & T5.0$\pm$0.5  &$-$0.57$\pm$0.03 (1)&$+$0.27$\pm$0.09&$-$0.27$\pm$0.16&$-$0.01$\pm$0.15& 0.34$\pm$0.06  &   84$\pm$10  & 1 \\
2M2331-4718 &2MASS J23312378-4718274   & T5        & T4.7$\pm$0.5  &$-$0.51$\pm$0.03 (1)&$+$0.15$\pm$0.16&$+$0.12$\pm$0.25&$+$0.27$\pm$0.21& 0.20$\pm$0.07  &  118$\pm$6   & 1 \\[2pt]
{\bf Other} & \\
SD0151+1244 &SDSSp J015141.69+124429.6 & T1        & L5-T0         &$+$0.18$\pm$0.04 (2)&$+$0.96$\pm$0.17&$+$0.42$\pm$0.22&$+$1.38$\pm$0.23& 0.90$\pm$0.05  &  100$\pm$2   & 1 \\
SD0207+0000 &SDSSp J020742.83+000056.2 & T4        & T4.0$\pm$0.8  &$-$0.37$\pm$0.10 (1)&$+$0.40$\pm$0.15&\nodata        :&$+$1.39$\pm$0.15:& 0.30$\pm$0.13  &  101$\pm$13  & 1 \\
2M0243-2453 &2MASS J02431371-2453298   & T6        & T5.8$\pm$0.5  &$-$0.77$\pm$0.02 (3)&$+$0.24$\pm$0.12&$-$0.08$\pm$0.20&$+$0.17$\pm$0.17& 0.28$\pm$0.05  &  229$\pm$25  & 1 \\
2M0415-0935 &2MASS J04151954-0935066   & T8        & T8.0$\pm$0.2  &$-$1.66$\pm$0.03 (3)&$+$0.16$\pm$0.13&$+$0.11$\pm$0.23&$+$0.27$\pm$0.21& 2.31$\pm$0.04  &   76$\pm$1   & 1 \\
SD0423-0414 &SDSSp J042348.57-041403.5 & T0        & L3-T0         &$+$0.15$\pm$0.02 (4)&$+$1.00$\pm$0.04&$+$0.53$\pm$0.05&$+$1.54$\pm$0.04& 0.29$\pm$0.04  &  281$\pm$9   & 1 \\
2M0516-0445 &2MASS J05160945-0445499   & T5.5      & T5.6$\pm$0.3  &$-$0.70$\pm$0.02 (3)&$+$0.26$\pm$0.18&$+$0.23$\pm$0.26&$+$0.50$\pm$0.22& 0.30$\pm$0.04  &  234$\pm$9  & 2 \\
2M0559-1404 &2MASS J05591914-1404488   & T4.5      & T4.1$\pm$0.4  &$-$0.40$\pm$0.02 (3)&$+$0.12$\pm$0.05&$+$0.10$\pm$0.07&$+$0.22$\pm$0.06& 0.72$\pm$0.04  &  120$\pm$1   & 1 \\
SD1021-0304 &SDSSp J102109.69-030420.1 & T3.5      & T2.3$\pm$1.0  &$-$0.11$\pm$0.04 (4)&$+$0.91$\pm$0.14&$+$0.22$\pm$0.20&$+$1.13$\pm$0.19& 0.17$\pm$0.04  &  252$\pm$40  & 1 \\
SD1110+0116 &SDSSp J111010.01+011613.1 & T5.5      & T5.5$\pm$0.5  &$-$0.66$\pm$0.12 (2)&$+$0.42$\pm$0.18&$+$0.79$\pm$0.14&$+$1.21$\pm$0.12:& 0.34$\pm$0.10  &  110$\pm$27  & 1 \\
SD1207+0244 &SDSSp J120747.17+024424.8 & T0.5      & T0  $\pm$1.5  &$+$0.05$\pm$0.03 (1)&$+$1.02$\pm$0.09&$+$0.57$\pm$0.09&$+$1.59$\pm$0.09& 0.39$\pm$0.09  &  286$\pm$13  & 1 \\
2M1217-0311 &2MASS J12171110-0311131   & T7        & T7.5$\pm$0.3  &$-$1.42$\pm$0.08 (1)&$+$0.11$\pm$0.13&$-$0.14$\pm$0.12&$-$0.03$\pm$0.06:& 1.00$\pm$0.06  &  278$\pm$3   & 1 \\
2M1225-2739 &2MASS J12255432-2739476AB & T6        & T6.4$\pm$0.3  &$-$0.89$\pm$0.06 (2)&$+$0.16$\pm$0.09&$+$0.03$\pm$0.17&$+$0.19$\pm$0.15& 0.72$\pm$0.03  &  146$\pm$1 & 1 \\
SD1254-0122 &SDSSp J125453.90-012247.4 & T2        & T0.1$\pm$1.5  &$+$0.04$\pm$0.02 (1)&$+$0.80$\pm$0.04&$+$0.25$\pm$0.06&$+$1.05$\pm$0.06& 0.45$\pm$0.06  &  285$\pm$6   & 1 \\
SD1346-0031 &SDSSp J134646.45-003150.4 & T6.5      & T5.8$\pm$0.4  &$-$0.87$\pm$0.08 (1)&$+$0.54$\pm$0.16&$-$0.31$\pm$0.30&$+$0.23$\pm$0.29& $<0.24$        &      \nodata     & 1 \\
Gl570d      &                          & T7.5      & T7.5$\pm$0.2  &$-$1.38$\pm$0.05 (4)&$+$0.06$\pm$0.10&$+$0.03$\pm$0.18&$+$0.08$\pm$0.16& 1.99$\pm$0.03  &  147.7$\pm$0.6 & 1 \\
2M1534-2952 &2MASS J15344984-2952274AB &\nodata    & T4.4$\pm$0.5  &$-$0.45$\pm$0.01 (1)&$+$0.03$\pm$0.11&$+$0.02$\pm$0.15&$+$0.06$\pm$0.13& 0.28$\pm$0.03  &  156$\pm$3   &   \\
2M1546-3325 &2MASS J15462718-3325111   &\nodata    & T5.1$\pm$0.4  &$-$0.59$\pm$0.01 (1)&$+$0.19$\pm$0.10&$-$0.04$\pm$0.20&$+$0.15$\pm$0.19& 0.25$\pm$0.04  &   38$\pm$12  &   \\
SD1624+0029 &SDSSp J162414.37+002915.6 & T6        & T5.9$\pm$0.4  &$-$0.79$\pm$0.03 (1)&$-$0.03$\pm$0.11&$+$0.01$\pm$0.10&$-$0.02$\pm$0.05:& 0.34$\pm$0.07  &  256$\pm$18  & 1 \\
SD1750+1759 &SDSSp J175032.96+175903.9 & T3        & T2.6$\pm$0.9  &$-$0.18$\pm$0.04 (1)&$+$0.39$\pm$0.16&$+$0.47$\pm$0.23&$+$0.86$\pm$0.21& 0.22$\pm$0.08  &  100$\pm$12  & 1 \\
$\epsilon$ Ind BC&                          &\nodata    & T1.7$\pm$1.0  &$-$0.09$\pm$0.01 (3)&$+$0.60$\pm$0.03&$+$0.10$\pm$0.03&$+$0.70$\pm$0.03& 4.74$\pm$0.02  &  120.6$\pm$0.2 &   \\
2M2228-4310 &2MASS J22282889-4310262   & T6        & T6.0$\pm$0.4  &$-$0.82$\pm$0.02 (1)&$+$0.30$\pm$0.14&$+$0.07$\pm$0.24&$+$0.37$\pm$0.22& 0.35$\pm$0.05  &  164$\pm$5   & 1 \\
2M2339+1352 &2MASS J23391025+1352284   & T5        & T4.6$\pm$0.4  &$-$0.47$\pm$0.03 (2)&$+$0.42$\pm$0.19&$-$0.32$\pm$0.34&$+$0.09$\pm$0.33& 1.12$\pm$0.20  &  156$\pm$4   & 1 \\
2M2356-1553 &2MASS J23565477-1553111   & T5.5      & T5.7$\pm$0.3  &$-$0.71$\pm$0.04 (4)&$+$0.19$\pm$0.11&$-$0.14$\pm$0.21&$+$0.05$\pm$0.19& 0.77$\pm$0.04  &  211$\pm$7   & 1 \\
\enddata                
\tablenotetext{a}{ Spectroscopic spectral type.}
\tablenotetext{b}{ Methane spectral type, derived from \Msl$_\mathrm{Diff}$ and equation \ref{methane_to_spectraltype}.}
\tablenotetext{c}{ See Section \ref{differential} for definition of \Msl$_\mathrm{Diff}$. The number of observations averaged to produce the measurement listed are indicated in parentheses.}
\tablenotetext{d}{ J-H, H-Ks and J-Ks on the 2MASS photometric system. ':' indicates object was not detected in the 2MASS Ks band,
                   and 95\% confidence upper limit has been used to derive the colour, the quoted uncertainty of which is due only to the
                   J detection. SD0207+0000 was not detected in H or Ks by 2MASS.}
\tablenotetext{e}{ Proper motion magnitude ($\mu$ in \arcsec/yr) and direction ($\theta$ in degrees east of 
                   north). Many of the T dwarfs listed in Table \ref{diff_table} will be at distances of
                   7-20\,pc, and will have annual parallax  motions of 0.125-0.05\arcsec, which we have not
                   attempted to control. These will introduce proper motion uncertainties of a magnitude
                   similar to those due to the precision of the astrometric transformation between the two
                   epochs}
\tablenotetext{f}{ Source for spectroscopic spectral type.  
                   1 - \citet{bu2005}. 
                   2 - spectra from \citet{le2002,kn2004,bu2003b,bu2003c,bu2004a} classified using procedure
                   described in Section \ref{spectral_classification}. 
                   3 - IRIS2 spectral types as described in Section \ref{spectral_classification}.
                   If no spectral type listed for a T dwarf then no ``hybrid''
                   spectral type has been published, or could be calculated from available archival data.}
\end{deluxetable}
\clearpage
\end{landscape}

\label{lastpage}

\end{document}